\documentclass{mn2e}
\usepackage[totalwidth=480pt, totalheight=680pt]{geometry}
\usepackage{epsfig}
\usepackage{rotating}
\usepackage{longtable}
\usepackage{graphics}
\usepackage{amssymb}
\usepackage{supertabular}
\usepackage{times}

\title[Deep survey of heavy-element lines in PNe -- I]
{A deep survey of heavy element lines in Planetary Nebulae --
I. Observations and forbidden-line densities, temperatures and
abundances}
\author[Y. G. Tsamis et al.]
{Y. G. Tsamis$^{1,2}$, M. J. Barlow$^1$, X.-W. Liu$^{1,3}$, I. J. Danziger$^4$
and P. J. Storey$^1$ \\
$^1$Department of Physics and Astronomy, University College London,
      Gower Street, London WC1E 6BT, UK\\
$^2$Met.\,Office, London Road, Bracknell RG12 2SZ, UK\\
$^3$Department of Astronomy, Peking University,
   Beijing 100871, P.R. China \\
$^4$Osservatorio Astronomico di Trieste, Via G. B. Tiepolo 11, I-34131 Trieste,
Italy}

\date{Received:}
\newcommand{\eld}{$N_{\rm e}$}
\newcommand{\crd}{$N_{\rm cr}$}
\newcommand{\elt}{$T_{\rm e}$}
\newcommand{\exe}{$E_{\rm ex}$}
\newcommand{\cmt}{cm$^{-3}$}
\newcommand{\cp}{C$^+$}
\newcommand{\cpp}{C$^{2+}$}
\newcommand{\cppp}{C$^{3+}$}
\newcommand{\cfp}{C$^{4+}$}
\newcommand{\op}{O$^+$}
\newcommand{\opp}{O$^{2+}$}
\newcommand{\oppp}{O$^{3+}$}
\newcommand{\ofp}{O$^{4+}$}
\newcommand{\np}{N$^+$}
\newcommand{\npp}{N$^{2+}$}
\newcommand{\nppp}{N$^{3+}$}
\newcommand{\nfp}{N$^{4+}$}
\newcommand{\nepp}{Ne$^{2+}$}
\newcommand{\neppp}{Ne$^{3+}$}
\newcommand{\nefp}{Ne$^{4+}$}
\newcommand{\Hb}{H$\beta$}
\newcommand{\foiii}{[O~{\sc iii}]}
\newcommand{\fniii}{[N~{\sc iii}]}
\newcommand{\foii}{[O~{\sc ii}]}
\newcommand{\fsii}{[S~{\sc ii}]}
\newcommand{\fsiii}{[S~{\sc iii}]}
\newcommand{\fnii}{[N~{\sc ii}]}
\newcommand{\fariv}{[Ar~{\sc iv}]}
\newcommand{\fcliii}{[Cl~{\sc iii}]}
\newcommand{\fneiii}{[Ne~{\sc iii}]}
\newcommand{\fneiv}{[Ne~{\sc iv}]}
\newcommand{\fnev}{[Ne~{\sc v}]}
\newcommand{\oiii}{O~{\sc iii}}
\newcommand{\sii}{S~{\sc ii}}
\newcommand{\nii}{N~{\sc ii}}
\newcommand{\niii}{N~{\sc iii}}

\newcommand{\oii}{O~{\sc ii}}
\newcommand{\cii}{C~{\sc ii}}
\newcommand{\neii}{Ne~{\sc ii}}
\newcommand{\ciii}{C~{\sc iii}}
\newcommand{\civ}{C~{\sc iv}}

\newcommand{\cliii}{Cl~{\sc iii}}

\newcommand{\ariv}{Ar~{\sc iv}}
\newcommand{\hi}{H\,{\sc i}}
\newcommand{\hii}{H~{\sc ii}}
\newcommand{\hei}{He~{\sc i}}
\newcommand{\heii}{He~{\sc ii}}

\newcommand{\hp}{H$^+$}

\begin{document}
\maketitle

\begin{abstract} \noindent We present deep optical spectrophotometry of
twelve Galactic planetary nebulae (PNe) and three Magellanic Cloud PNe. Nine of
the Galactic PNe were observed by scanning the spectrograph's slit across the
nebula, yielding relative line intensities for the entire nebula that are
suitable for comparison with integrated nebular fluxes measured in other
wavelength regions. In this paper we use the fluxes of collisionally excited
lines (CELs) from the nebulae to derive electron densities and temperatures,
and ionic abundances. We find that the nebular electron densities derived from
optical CEL ratios are systematically higher than those derived from the ratios
of the infrared (IR) fine-structure (FS) lines of {\foiii}. The latter have
lower critical densities than the typical nebular electron densities derived
from optical CELs, indicating the presence of significant density variations
within the nebulae, with the infrared CELs being biased towards lower density
regions.

We find that for several nebulae the electron temperatures obtained from
{\foii} and {\fnii} optical CELs are significantly affected by recombination
excitation of one or more of the CELs. When allowance is made for recombination
excitation, much better agreement is obtained with the electron temperatures
obtained from optical {\foiii} lines. We also compare electron temperatures
obtained from the ratio of optical nebular to auroral {\foiii} lines with
temperatures obtained from the ratio of {\foiii} optical lines to {\foiii} IR
FS lines. We find that when the latter are derived using electron densities
based on the {\foiii} 52\,$\mu$m/88\,$\mu$m line ratio, they yield values that
are significantly higher than the optical {\foiii} electron temperatures. In
contrast to this, {\foiii} optical/IR temperatures derived using the higher
electron densities obtained from optical {\fcliii} $\lambda$5517/$\lambda$5537
ratios show much closer agreement with optical {\foiii} electron temperatures,
implying that the observed {\foiii} optical/IR ratios are significantly
weighted by densities in excess of the critical densities of both {\foiii} FS
lines. Consistent with this, ionic abundances derived from {\foiii} and
{\fniii} FS lines using electron densities from optical CELs show much better
agreement with abundances derived for the same ions from optical and
ultraviolet CELs than do abundances derived from the FS lines using the lower
electron densities obtained from the observed {\foiii} 52\,$\mu$m/88\,$\mu$m
ratios. The behaviour of electron temperatures obtained making use of the
temperature-insensitive {\foiii} IR FS lines provides no support for
significant temperature fluctuations within the nebulae that could be
responsible for derived Balmer jump electron temperatures being lower than
temperatures obtained from the much more temperature sensitive {\foiii} optical
lines.

\end{abstract}

\begin{keywords}
ISM: abundances -- ISM: lines -- planetary nebulae: general
\end{keywords}

\section{Introduction}

This is the first of two papers devoted to the study of elemental abundances in
a sample of Galactic and Magellanic Cloud planetary nebulae which, together
with a companion paper by Tsamis et al. (2003c; Paper II), focus on the problem
of the optical recombination-line emission from heavy element ions (e.g.
{\cpp}, {\npp}, {\opp}) in planetary nebulae (PNe) and {\hii} regions. The main
manifestation of this intriguing question is the observed discrepancy between
nebular elemental abundances derived from weak, optical recombination lines
(ORLs; such as {\cii} $\lambda$4267, {\nii} $\lambda$4041, {\oii}
$\lambda\lambda$4089, 4650) on the one hand and the much brighter
collisionally-excited lines (CELs; often collectively referred to as forbidden
lines) on the other (Liu et al. 1995, 2000, 2001b; Tsamis 2002; Tsamis et al.
2003a), with ORLs typically being found to yield ionic abundances that are
factors of two or more larger than those obtained from CELs emitted by the same
ions. A closely linked problem involves the observed disparity between the
nebular electron temperatures derived from the traditional {\foiii}
($\lambda$4959 + $\lambda$5007)/$\lambda$4363 ratio and the {\hi} Balmer
discontinuity diagnostic: the latter yields temperatures that are in most cases
lower than those derived from the {\foiii} ratio (Peimbert 1971; Liu \&
Danziger 1993; Liu et al. 2001b; Tsamis 2002).

Tsamis et al. (2003a) have presented new spectroscopic observations and an
abundance analysis of the Galactic {\hii} regions M\,17 and NGC\,3576 and the
Magellanic Cloud {\hii} regions 30~Doradus, LMC~N11B and SMC~N66, involving
both CELs and ORLs. We reported surprisingly large ORL vs. CEL abundance
discrepancies in these objects: the {\opp}/{\hp} abundance ratios derived from
multiple {\oii} ORLs were found to be higher than the corresponding values
derived from {\foiii} CELs, typically by a factor of $\sim$\,2, though for one
nebula, LMC~N11B, a factor of five discrepancy was found.

In the present paper, we present deep, long-slit, optical spectrophotometry of
12 Galactic and 3 Magellanic Cloud PNe. We present an extensive list of all the
detected emission lines, including the numerous {\cii}, {\nii}, {\oii} and
{\neii} ORLs, along with their fluxes and dereddened intensities. We undertake
a \emph{forbidden-line} abundance study deriving C, N, O, Ne, S, Cl and Ar
abundances from CELs, as well as He abundances from {\hei} and {\heii} ORLs.
Along with the optical data, we analyze \emph{International Ultraviolet
Explorer} NEWSIPS spectra of 10 PNe, which yield fluxes of important C, N, O
and Ne CELs. We further derive abundances for a number of PNe from the far-IR
{\foiii} 52- and 88-$\mu$m and {\fniii} 57-$\mu$m fine-structure lines using
the \emph{Infrared Space Observatory} LWS line fluxes by Liu et al. (2001a), as
well as from the \emph{IRAS} {\fneiii} 15.5-$\mu$m line fluxes by Pottasch et
al. (1984). In an accompanying paper (Tsamis et al., Paper~II), heavy-element
ionic abundances are derived from the observed {\cii}, {\ciii}, {\civ}, {\nii},
{\niii}, {\oii} and {\neii} ORLs and an extensive comparison between ORL and
CEL elemental abundance ratios is made.

In Section~2 we describe our optical spectroscopic observations obtained with
the ESO 1.52-m and NTT 3.5-m telescopes, as well as the data reduction process,
and present emission line fluxes for all the PNe; we also present \emph{IUE}
ultraviolet (UV) observations and our data analysis methods. In Section~3 we
describe the corrections for interstellar extinction to the UV and optical
data. In Section~4 we present derived nebular electron temperatures and
densities. In Section~5 we present our derived He/H ratios and the heavy
element abundances derived from UV, optical and infrared (IR)
collisionally-excited lines.

\section{Observations}

\subsection{Optical spectroscopy}

\subsubsection{Description and data reduction}

The planetary nebulae studied in this work were observed at the European
Southern Observatory (ESO), using the 1.52-m telescope for the Galactic objects
and the 3.5-m New Technology Telescope (NTT) for the Magellanic Cloud objects.
The long-slit observations were taken during July and December 1995, July 1996
and February 1997. An observational journal is presented in Table~1.

The Boller \& Chivens (B \& C) spectrograph was used on the 1.52-m telescope in
July 1995, equipped with a Ford $2048\,\times\,2048$ pixel CCD detector,
with each pixel 15$\mu$m\,$\times$\,15$\mu$m in size. This detector was superseded
in 1996 and 1997 by a UV-enhanced Loral $2048\,\times\,2048$,
15$\mu$m\,$\times$\,15$\mu$m CCD of improved quantum efficiency. A 2\,arcsec
wide, 3.5\,arcmin long-slit was employed. During all runs the CCDs were binned
by a factor of 2 along the slit direction, in order to reduce the read-out
noise, yielding a spatial sampling of 1.63\,arcsec per pixel projected on the
sky. A 2400 lines per mm grating was used in first order, along with an
order-sorting WG360 filter, to cover the 3995--4978\,{\AA} wavelength range at
a spectral resolution of 1.5\,{\AA} FWHM. A second grating in first order,
along with a WG345 filter, was used to cover the 3535--7400\,{\AA} range at a
resolution of 4.5\,{\AA} (excluding NGC\,2440 for which 4.5\,{\AA} resolution
spectra were not obtained). For three objects (NGC\,5882, NGC\,6302 and NGC\,6818)
an extra wavelength range (3040--4040\,{\AA}) was covered with the holographic
2400 lines\,per\,mm grating at a resolution of 1.5\,{\AA}, so that the
high-order H\,{\sc i} Balmer lines---transitions \emph{n}\,$\rightarrow$\,2,
from upper levels of principal quantum number up to \emph{n} = 24---were
accurately recorded. The low resolution (4.5\,{\AA}), wider spectral coverage
spectra, yielded fluxes of forbidden lines of N$^+$, O$^+$, O$^{2+}$,
Ne$^{2+}$, S$^+$, S$^{2+}$, Cl$^{2+}$, Ar$^{2+}$ and Ar$^{3+}$ etc. ions, as
well as H$^0$, He$^0$ and He$^+$ lines, and covered the Balmer discontinuity
(or Balmer jump; BJ). The higher resolution (1.5\,{\AA}), deep spectra recorded
the optical recombination lines of C~{\sc ii}, C~{\sc iii}, {\civ}, N~{\sc ii},
N~{\sc iii}, {\oii} and Ne~{\sc ii}, the majority of which lay between
4000--5000\,{\AA}.

Typical exposure times for these Galactic PNe ranged between 20\,sec and
35\,min. The short exposures were carefully chosen so that strong nebular
lines, such as H$\alpha$, H$\beta$ or the [O~{\sc iii}] $\lambda\lambda$4959,
5007 and [N~{\sc ii}] $\lambda\lambda$6548, 6584 lines, would not be saturated.
On the other hand, the deep exposures were aimed at capturing the faint ORLs,
whose typical intensities can be as low as $10^{-4}$ that of H$\beta$, at an
appropriate resolving power.

For nine out of twelve observed Galactic nebulae (NGC entries 2022, 2440, 3132,
3242, 3918, 5315, 5882 and IC\,4191 and 4406) mean spectra were obtained by
uniformly scanning the long-slit of the B\&C spectrograph across the nebular
surfaces (IC\,4406 was scanned only at a lower resolution of 4.5\,{\AA}). For
these nebulae total fluxes of all detected lines can be derived, when total
nebular H$\beta$ line fluxes are adopted from the compilation of Cahn, Kaler \&
Stanghellini (1992; hereafter CKS92), as listed in column~2 of Table~1.
Furthermore, these mean spectra are directly comparable with UV and IR spectra
obtained from space-borne facilities (such as the \emph{IUE}, \emph{IRAS} and
\emph{ISO} satellites), whose large apertures usually capture most or all of
the nebular emission, depending on the angular size of the observed nebulae and
the pointing coordinates of such observations. In this way, several ionization
stages of the same element can be traced, allowing a more complete picture to
be drawn for properties such as reddening, excitation class and elemental
abundances.

For the remaining three Galactic PNe, plus IC\,4406 in the 3995--4978\,{\AA}
range only, fixed-slit spectra were obtained (Table~1). In those cases, the
slit was positioned through the nebular center -- using the central star of the
PN as a guide, if visible -- and passing through the visually brightest parts
of the nebula.

The Magellanic Cloud planetary nebulae LMC~N66, LMC~N141 and SMC~N87 were
observed at ESO with the 3.5-m NTT, in December 1995 (Table~1). The ESO
Multi-Mode Instrument (EMMI) was used in the following modes: red imaging and
low dispersion grism spectroscopy (RILD), blue medium dispersion spectroscopy
(BLMD) and dichroic medium dispersion spectroscopy (DIMD). The detector was a
TEK $1024\,\times\,1024$, 24$\mu$m\,$\times$\,24$\mu$m CCD (no. 31; on the blue
arm), used while in BLMD observing mode and a TEK $2048\,\times\,2048$,
24$\mu$m\,$\times$\,24$\mu$m CCD (no. 36; on the red arm), while in RILD mode.
Both cameras were in use when observing in DIMD mode. In those cases, a dichroic
prism is inserted into the beam path so that light is directed to the blue and
red grating units in synchronization, allowing simultaneous exposures to be
obtained in the blue and red parts of the optical spectrum. For all exposures,
both CCDs were binned by a factor of 2 in both directions, in order to reduce
the read-out noise. The spatial sampling was thus 0.74 and 0.54\,arcsec per
pixel projected on the sky, for CCDs no. 31 and no. 36, respectively.

Four wavelength regions were observed with two different gratings (\#3, \#7) at
spectral resolutions of approximately 2\,\AA~FWHM (3635--4145, 4060--4520,
4515--4975\,{\AA}), and 3.8\,\AA~(6507--7828\,{\AA}), respectively. An extra
wavelength range, 3800--8400\,{\AA}, was covered only for LMC~N66 at a
resolution of 11\,{\AA}, using a grism (\#3). An OG530 filter was used when
observing in DIMD mode. The slits used were 5.6\,arcmin long and 1 and
1.5\,arcsec wide, but wide-slit 5\,arcsec grism observations were also taken
for LMC~N66. Exposure times for the Magellanic Cloud PNe ranged from 3 to
30\,min.

\setcounter{table}{0}
\begin{table*}
\centering
\begin{minipage}{120mm}
\caption{Journal of ESO optical observations.}
\begin{tabular}{lclrcccc}
\noalign{\vskip3pt}\noalign{\hrule}\noalign{\vskip3pt}
PN&log$F$(H$\beta$) & $c$(H$\beta)^{\rm radio}$   &Date        &$\lambda$-range        &FWHM           &Mode &Exp. Time \\
  & (ergs~cm$^{-2}$~s$^{-1}$) &   &(UT)        &({\AA})                  &({\AA})          &
&(sec)             \\
\noalign{\vskip3pt} \noalign{\hrule} \noalign{\vskip3pt}

\multicolumn{6}{c}{\underline{ESO 1.52-m}} \\
\noalign{\vskip2pt}
NGC\,2022&--11.13&0.42  &7 Feb 1997  &3530--7428             &4.5            &scanning   &40, 1155 \\
     & &          &8,\,10 Feb 1997  &3994--4978        &1.5            &"          &3\,$\times$\,1500, 1800, 2100 \\
\noalign{\vskip2pt}
NGC\,2440&--10.50&0.47 &9 Feb 1997  &3994--4978             &1.5            &scanning   &1800 \\
\noalign{\vskip2pt}
NGC\,3132&--10.45&0.30 &7 Feb 1997  &3530--7428             &4.5            &scanning   &70, 1147 \\
     & &   &8,\,10 Feb 1997  &3994--4978        &1.5            &"          &2\,$\times$\,1620, 2\,$\times$\,1800\\
\noalign{\vskip2pt}

NGC\,3242&--9.79&0.17 &7 Feb 1997  &3530--7428             &4.5            &scanning   &25, 480  \\
          & &  &8,\,10 Feb 1997  &3994--4978        &1.5            &"          &1614, 2\,$\times$\,1800, 2100 \\
\noalign{\vskip2pt}
NGC\,3918&--10.04&0.40 &7 Feb 1997  &3530--7428             &4.5            &scanning   &20, 22, 40, 200 \\
      & &   &8,\,10 Feb 1997  &3994--4978        &1.5            &"          &1020, 1200, 2\,$\times$\,1620 \\
\noalign{\vskip2pt}
NGC\,5315&--10.42&0.55 &7 Feb 1997  &3530--7428             &4.5            &scanning   &30, 240 \\
        & &  &9,\,10 Feb 1997  &3994--4978        &1.5            &"          &1080, 1350, 2\,$\times$\,1500 \\
\noalign{\vskip2pt}
NGC\,5882&--10.38&0.42 &12 Jul 1996 &3040--4050             &1.5            &fixed slit &2\,$\times$\,1800 \\
        & & &7 Feb 1997  &3530--7428             &4.5            &scanning   &20, 220 \\
        & & &9,\,10 Feb 1997 &3994--4978         &1.5            &"          &3\,$\times$\,1500, 1800 \\
\noalign{\vskip2pt}
NGC\,6302&--10.55&1.39 &11 Jul 1996 &3040--4050             &1.5            &fixed slit &2\,$\times$\,1800 \\
         & &  &29 Jul 1995 &3530--7428             &4.5            &"          &30, 60, 300 \\
         & &  &23 Jul 1995 &3994--4978             &1.5            &"          &3\,$\times$\,1800 \\
\noalign{\vskip2pt}
NGC\,6818&--10.48&0.37 &11 Jul 1996 &3040--4050             &1.5            &fixed slit &2\,$\times$\,1800 \\
         & &  &29 Jul 1995 &3530--7428             &4.5            &"          &60, 300 \\
         & &  &23 Jul 1995 &3994--4978             &1.5            &"          &3\,$\times$\,1800 \\
\noalign{\vskip2pt}
IC\,4191&--10.99&0.70 &7 Feb 1997  &3530--7428             &4.5            &scanning   &2\,$\times$\,30, 500 \\
        & &  &8 Feb 1997  &3994--4978             &1.5            &"          &900, 1200 \\
        & &  &29 Jul 1995 &3530--7428             &4.5            &fixed slit &5, 10, 300, 600 \\
        & &  &22 Jul 1995 &3994--4978             &1.5            &"          &3\,$\times$\,1800 \\
\noalign{\vskip2pt}
IC\,4406&--10.75&0.27 &7 Feb 1997  &3530--7428             &4.5            &scanning   &50, 1080\\
        & &  &29 Jul 1995 &3530--7428             &4.5            &fixed slit &2\,$\times$\,300\\
        & &  &22 Jul 1995 &3994--4978             &1.5            &"          &5\,$\times$\,1800 \\
\noalign{\vskip2pt}
My\,Cn\,18&--11.21&0.74 &29 Jul 1995 &3530--7428             &4.5            &fixed slit &2\,$\times$\,120, 600\\
          & & &23 Jul 1995 &3994--4978             &1.5            &"          &4\,$\times$\,1800 \\
\noalign{\vskip3pt}
\multicolumn{6}{c}{\underline{NTT 3.5-m}} \\
\noalign{\vskip2pt}
SMC N87&--12.48&  &17 Dec 1995 &6507--7828             &3.8            &fixed slit &1200 \\
       & &   &"           &3635--4145             &2              &"          &1200 \\
       & &   &"           &4060--4520             &2              &"          &2\,$\times$\,1800 \\
       & &   &"           &4515--4975             &2              &"          &900, 3\,$\times$\,1800 \\
\noalign{\vskip2pt}
LMC N66&--12.68& &17 Dec 1995 &3800--8400              &11             &fixed slit  &2\,$\times$\,300 \\
       & &   &"           &6507--7828             &3.8            &"          &300  \\
       & &   &"           &3635--4145             &2              &"          &300 \\
       & &   &"           &4060--4520             &2              &"          &360 \\
       & &   &"           &4515--4975             &2              &"          &600 \\
\noalign{\vskip2pt}
LMC N141&--12.48& &17 Dec 1995 &6507--7828             &3.8            &fixed slit  &1200 \\
        & &   &"           &3635--4145             &2              &"          &1800 \\
        & &   &"           &4060--4520             &2              &"          &2\,$\times$\,1800 \\
        & &   &"           &4515--4975             &2              &"          &180, 4\,$\times$\,1800 \\
\noalign{\vskip3pt} \noalign{\hrule} \noalign{\vskip3pt}
\end{tabular}
\end{minipage}
\end{table*}
\noindent

The two-dimensional Galactic PN spectra were reduced with the {\sc long92}
package within {\sc midas}. They were bias-subtracted, flat-fielded via
division by a normalized flat field, cleaned of cosmic-rays, and then
wavelength-calibrated using exposures of a He-Ar calibration lamp. The effect
of atmospheric extinction was removed by correcting the spectra using
extinction coefficients for the La Silla site. During the 1997 run, twilight
sky flat-fields were also obtained with the purpose to correct the small
variations in illumination along the slit, which were $\sim$\,1--3 per cent.
For the 1995 and 1996 runs, the spectra were flux-calibrated using wide-slit
(8\,arcsec) observations of the \emph{HST} standard stars Feige\,110 and the
nucleus of the PN NGC\,7293 (Walsh 1993). The flux distributions of the
standards were modeled with high-order spline fits. In 1997, the CTIO standard
stars LTT\,4363, LTT\,6248 (Hamuy et al. 1994) and the \emph{HST} standard
HD\,49798 (Walsh 1993) were used.

The 3040--4040\,{\AA} spectra of NGC\,5882, 6302 and 6818 are affected by ozone
absorption bands, shortwards of 3400\,{\AA} (Schachter 1991). The bands have a
typical width of $\sim$\,15\,{\AA}, i.e. much wider than the nebular emission
lines ($\sim$\,1.5\,{\AA}). In order to remove the ozone absorption we used the
ozone-opacity template derived by Liu et al. (2000) for the study of NGC\,6153
from the same site (ESO, La Silla) as the observations presented herein. The
standard stars Feige\,110 and the nucleus of NGC\,7293 were observed with a
narrow 2\,arcsec slit, i.e. the same as the one used for nebular observations.
These narrow-slit spectra were used to derive the ozone opacity per unit
airmass as a function of wavelength, relative to the mean atmospheric
extinction curve at La Silla. The opacity curve, scaled by the airmasses of the
nebular exposures, was then used to divide out the ozone absorption bands.

The Cloud PN two-dimensional spectra were reduced with similar procedures as
above; the wavelength calibration was done with respect to Th-Ar and He-Ar
calibration lamps within {\sc midas}, while flux calibration was performed with
respect to wide-slit (8\,arcsec) observations of the standard Feige~110, within
the {\sc iraf} package.

For all flux-calibrated spectra suitable sky windows were selected on each side
of the nebular emission, their increments were summed-up, scaled to the total
number of CCD pixels along the slit direction and subtracted. In this way we
were able to subtract the sky emission for all PNe, since their angular
diameters were always effectively contained within our long-slit. The resulting
spectra were subsequently averaged along the spatial nebular emission---defined
by the extent of the H$\beta$ line---to yield the one-dimensional spectra on
which our nebular analysis was performed. In a couple of cases, such as
NGC\,6302 and NGC\,6818 for which only fixed-slit spectra were obtained, the
ORLs detected were rather weak and the spectra slightly noisier than the norm,
so we decided to integrate the spectra along the spatial extent of the C~{\sc
ii} $\lambda$4267 line which is one of the strongest heavy element
recombination lines in nebulae (the H$\beta$ line was of course sampled over
the same pixel numbers). This resulted in better S/N ratios for the O~{\sc ii}
and N~{\sc ii} recombination lines too, which are expected to originate from
similar nebular volumes as the C~{\sc ii} $\lambda$4267 line.

In other objects it was found that light from the nebula's central star affects
considerably the nebular line fluxes or the shape of the observed continuum
spectrum. For instance, NGC\,3132 harbours a binary nucleus whose primary
component is the A-type star HD\,87892 (Kohoutek \& Laustsen 1977; Sahu \&
Desai 1986). The absorption line spectrum of this star affects severely the
nebular Balmer lines by diminishing their intensities, so a few central CCD
pixels had to be excluded from the integration in order to prevent such
contamination and to restore the nebular line profiles. Another example is
NGC\,5882 which possesses a weak emission line (`\emph{wel}'),
hydrogen-deficient nucleus, whose continuum energy distribution evidently
affects the shape of the nebular continuum in the region of the Balmer
discontinuity ($\sim$\,3646\,{\AA}). Thus in order to get an accurate
measurement of the Balmer jump intensity and the high-order Balmer lines from
the deep 3040--4040\,{\AA} spectra of this object, we again excluded several
central pixels when averaging the two-dimensional CCD frames. This restored the
nebular continuum to its expected shape at those wavelengths.

In Fig.~\,1 we show a representative high resolution optical spectrum of
NGC\,3242 taken in scanning mode. It is a time-weighted average of four deep
exposures which captured a particularly rich set of recombination lines from
heavy element ions. Indicative line identifications are shown; they can be
found in full in Table~3.

\setcounter{figure}{0}
\begin{figure*}
\begin{center} \epsfig{file=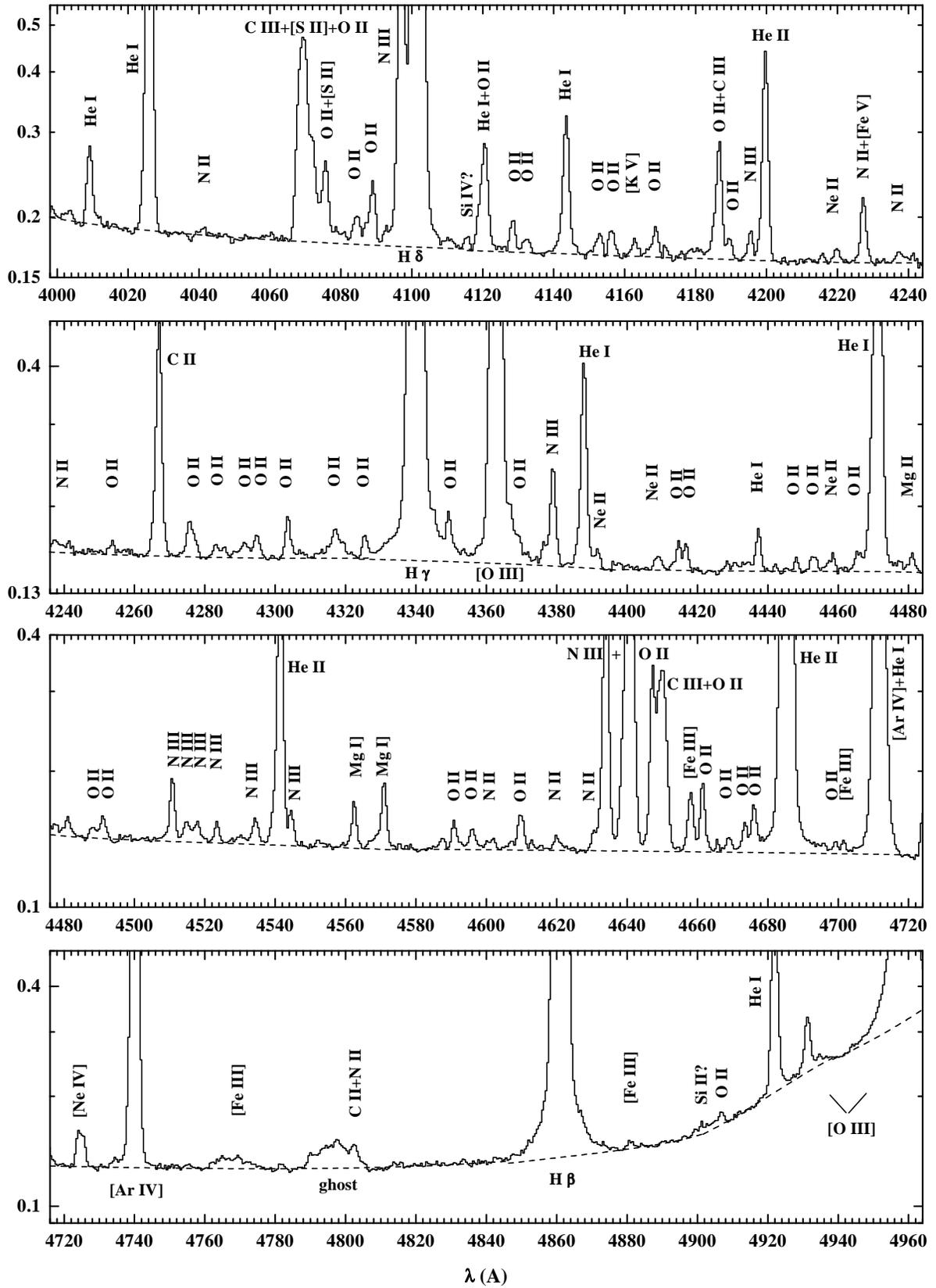, width=16.5 cm, clip=}
\caption{The spectrum of NGC\,3242 from 4000 to 4960\,{\AA} featuring the
prominent recombination lines from C, N, O and Ne ions; it was obtained by
uniformly scanning the entire nebular surface using a narrow long-slit. The
dashed lines show the adopted continuum level. The steep rise of the continuum
after 4880\,{\AA} is an instrumental artifact. The intensity is in units such
that $F$(H$\beta$)~=~100; the spectrum is not corrected for interstellar
extinction.}
\end{center}
\end{figure*}

\subsection{The \emph{IUE} observations}

We complemented our optical dataset with \emph{IUE} ultraviolet observations of
10 of the Galactic PNe, together with \emph{IUE} observations of the 3
Magellanic Cloud nebulae; the remaining two Galactic PNe, IC\,4191 and
My\,Cn\,18, were never observed by the \emph{IUE}. Low-resolution, large
aperture spectra obtained with the SWP and LWP cameras were accessed via the
Space Telescope Science Institute's website. The wavelength coverage of the two
cameras is 1150--1975, and 1910--3300\,{\AA}, respectively. All observations
were obtained with the \emph{IUE} large aperture, a $10.3\,\times\,23$
arcsec$^{2}$ oval. The data were retrieved in final (NEWSIPS) calibrated form.
When several spectra for one object were available, they were co-added weighted
by the integration time. The employed \emph{IUE} exposures are listed in
Table~2. For NGC\,3918 the line fluxes reported by Clegg et al. (1987) were
used, combined with those retrieved from a more recent exposure (SWP~47514).
Similarly, in the analysis of NGC\,5315 and NGC\,6818 we used the line fluxes
recently listed by Feibelman (1998; see his Table~3) and by Hyung, Aller \&
Feibelman (1999; see their Tables~4 and 5), respectively. Regarding LMC~N66, we
used the observed \emph{IUE} line fluxes listed by Pena \& Ruiz (1988; from
spectrum SWP~19905).

\setcounter{table}{1}
\begin{table}
\centering
\begin{minipage}{80mm}
\caption{Journal of \emph{IUE} observations.$^a$}
\begin{tabular}{llc}
\noalign{\vskip3pt}\noalign{\hrule}\noalign{\vskip3pt}
PN             &Spectrum                    &Exp. Time (sec)      \\
\noalign{\vskip3pt}\noalign{\hrule}\noalign{\vskip3pt}
NGC\,2022      &SWP 32732, 32729          &4800, 5400   \\
NGC\,2440      &SWP 07264, 14850, 17243  &1800, 2100, 360  \\
               &SWP 32720                  &2400            \\
NGC\,3132      &SWP 06160, 01745, 49629  &7200, 1200, 7200  \\
NGC\,3242      &SWP 5154, 5155            &300, 330   \\
NGC\,3918      &SWP 47514                  &300    \\
NGC\,5882      &SWP 16349                  &1800   \\
NGC\,6302      &SWP 05210, 05211          &720, 5400     \\
               &SWP 30986, 33379          &14400, 6600   \\
IC\,4406       &SWP 23420, 36237          &3600, 7200    \\
               &LWP 03725, 15490          &3600, 7200    \\
SMC~N87       &LWR 10043, SWP 13387       &10500, 13387  \\
LMC~N141      &SWP 13408                  &9000           \\
\noalign{\vskip3pt}\noalign{\hrule}\noalign{\vskip3pt}
\end{tabular}
\begin{description}
\item[$^a$]NGC\,5315 and NGC\,6818 line fluxes were taken from Feibelman
(1998) and Hyung et al. (1999) respectively; for NGC\,3918 the combined fluxes
from SWP~47514 and those of Clegg et al. (1987) were used.
\end{description}
\end{minipage}
\end{table}

The apparent angular sizes of all nebulae in our sample, apart from NGC\,5315
and the Magellanic Cloud nebulae, are larger than the \emph{IUE} large
aperture. Thus, in most cases only a certain fraction of the UV nebular
emission was captured by the satellite. This delicate aspect of the
observations is discussed in the following Section.

\subsection{Data analysis}

With respect to the optical data we proceeded by normalizing the individual
one-dimensional spectra to a flux scale such that \emph{F}(H$\beta$)~=~100, in
order to merge them afterwards. When H$\beta$ was found to be saturated in one
spectrum, the scaling was achieved using a suitable unsaturated H~{\sc i}
Balmer line from a shorter duration exposure. The 3040--4040\,{\AA} Galactic PN
spectra were brought to scale with the rest, using emission lines detected in
common with the 3535--7400\,{\AA} spectra, such as H\,9 $\lambda$3835, [Ne~{\sc
iii}] $\lambda$3868, H\,8 + He~{\sc i} $\lambda$3889, and [Ne~{\sc iii}]
$\lambda$3967 + H$\epsilon$ $\lambda$3970. In a similar fashion, the Magellanic
Cloud PN spectra from different grating settings were brought to scale via
their overlapping portions. For two of them, SMC~N87 and LMC~N141, their
6507--7828\,{\AA} spectra had no overlap in wavelength with other grating
settings. In this case we normalized the spectra to \emph{F}(H$\beta$)~=~100,
using the theoretical intensity ratio
\emph{I}(H$\alpha$)/\emph{I}(H$\beta$)~=~2.85, a value predicted for
representative nebular conditions of $T_{\rm e}$ = 10000\,K and $N_{\rm e}$ =
5000\,cm$^{-3}$ (Storey \& Hummer 1995); we also took into account the amount
of reddening estimated towards these objects (Section~5). Finally, the
normalized spectra taken at each grating setting were co-added, weighted by the
exposure time, in order to achieve an optimum S/N ratio.

Regarding the UV data, a way had to be found to scale them to the normalized
optical spectra. For all Galactic nebulae (except NGC\,5315, NGC\,6818 and
IC\,4406), we used the theoretical intensity ratio for He~{\sc ii}
$\lambda$1640/$\lambda$4686, calculated for the appropriate nebular
temperatures and densities using the recombination theory predictions of Storey
\& Hummer (1995). In this way we scaled the UV spectra to
\emph{F}(H$\beta$)~=~100 using our measurements of the dereddened
\emph{I}($\lambda$1640) and \emph{I}($\lambda$4686) fluxes. In cases where the
[O~{\sc ii}] $\lambda$2470 doublet line was detected, as in NGC\,6818 and
IC\,4406, we used instead the predicted intensity ratio [O~{\sc ii}]
($\lambda$7320 + $\lambda$7330)/$\lambda$2470~=~1.33, which holds independently
of nebular conditions since all lines involved arise from the same upper levels
(\emph{A}-values taken from Zeippen 1982). The method was found to be highly
reliable; for instance in NGC\,5315 whose angular diameter fits completely
within the \emph{IUE} large aperture---and there is thus no need for any
aperture correction to be applied---the [O~{\sc ii}] $\lambda$2470 intensity
predicted using the optical [O~{\sc ii}] $\lambda\lambda$7320, 7330
intensities, is within 2\,per cent of the actual UV measurement. For NGC\,5315
and the three Cloud PNe (SMC~N87, LMC~N66 and LMC~N141) the dereddened UV line
fluxes were scaled directly to \emph{F}(\Hb)~=~100, with the use of total {\Hb}
fluxes from CKS92 in the former case and from those listed by Barlow
(1987) in the latter cases.


We proceeded with measuring the emission line fluxes from the UV and optical
spectra using the {\sc midas} package. First we scrutinized the low resolution
\emph{IUE} spectra in order to determine the nature of the emission lines,
which in some cases can have a stellar and not nebular origin (e.g. N\,{\sc v}
$\lambda$1240, C~{\sc iv} $\lambda$1550, He~{\sc ii} $\lambda$1640). The FWHM
of the lines was compared against that of \emph{IUE} low resolution spectra
which is $\sim$6--7\,{\AA}. Lines with FWHM of $\gtrsim$\,8--9\,{\AA}were
checked for blending or for the possible presence of P\,Cygni features that
would betray their origin in a central star's wind, as is the case for instance
in NGC\,5315 and NGC\,5882.

Most of the line fluxes---and certainly all those of heavy element
recombination lines---were derived using Gaussian line profile fitting
techniques, apart from the strongest ones, for which their fluxes were measured
by simply integrating over the lines. Gaussian line fitting within {\sc midas}
is performed through least-squares fitting using the Newton-Raphson method.

Even in our 1.5\,{\AA} high-resolution optical spectra, blending of emission
lines is evident throughout the wavelength range covered and it particularly
affects weak features like the ORLs of interest. The line broadening is
dominated by instrumental broadening and not by nebular dynamics, since the
typical FWHM of 1.5\,{\AA} translates to $\sim$93 km\,s$^{-1}$ at 4861\,{\AA},
whereas typical PN expansion velocities are $\lesssim$25 km\,s$^{-1}$ (e.g.
Kwok 1994). Therefore, in order to de-convolve features affected by blending,
multiple Gaussian fitting was employed. In such cases a successful estimate of
the continuum emission level was deemed to be a very important first step.
After subtracting the local continuum, line fluxes were retrieved by fitting
multiple Gaussians of appropriate central wavelength and usually equal FWHM.
The FWHM was taken to be the same as that of nearby unblended lines of similar
strength to the ones fitted. Their relative wavelength spacing was constrained
to that from laboratory wavelengths. This procedure assured accurate flux
retrieval and aided line identification in the case of ambiguous features.

In rare instances where the fitting process was not adequately yielding unique
fluxes for closely spaced lines, we made use of the theoretical intensity
ratios of selected lines, assuming \emph{LS}-coupling to hold for the ions in
question (e.g. Kuhn 1969). As an example, in PNe of relatively high excitation
class, such as NGC\,5882, the [S~{\sc ii}] $\lambda$4068 line is blended with
the C~{\sc iii} V\,16 triplet, whose components at 4067.94, 4068.91, and
4070.26\,{\AA} have intensity ratios under pure \emph{LS}-coupling of 1.00 :
1.31 : 1.71 (e.g. Allen 1973). Also present in this blend are the O~{\sc ii}
V\,10 multiplet $\lambda\lambda$4069.62, .89 and 4072.16 lines. We therefore
fitted 6 Gaussians to the 4069\,{\AA} complex, fixing the C~{\sc iii} triplet
relative intensities according to theory, while allowing the intensities of the
other lines to vary independently. After reducing by 2 the number of free
parameters in the minimization problem, convergence was more readily achieved.
Of course in this way, the resulting flux for the C~{\sc iii} triplet refers to
the sum of its components only, while information on the true allocation of
flux among them is lost. Individual fluxes, however, are retrieved for the
[S~{\sc ii}] and O~{\sc ii} lines and the overall shape of the blend is
accurately reproduced (see Fig.\,~2). As another example, the blend at
4650\,{\AA} can be fitted with 5 Gaussians, including the O~{\sc ii}
$\lambda\lambda$4649.13, 4650.84 lines from multiplet V\,1 and the C~{\sc iii}
V\,1 triplet $\lambda\lambda$4647.42, 4650.25 and 4651.47 lines, the latter
having relative intensities of 5 : 3 : 1 as predicted under \emph{LS}-coupling.
In general, apart for the aforementioned cases we avoided fixing the relative
strengths of ORLs, since we were interested to compare our observations with
the predictions of current recombination theory, especially for O~{\sc ii}.

\setcounter{figure}{1}
\begin{figure}
\begin{center} \epsfig{file=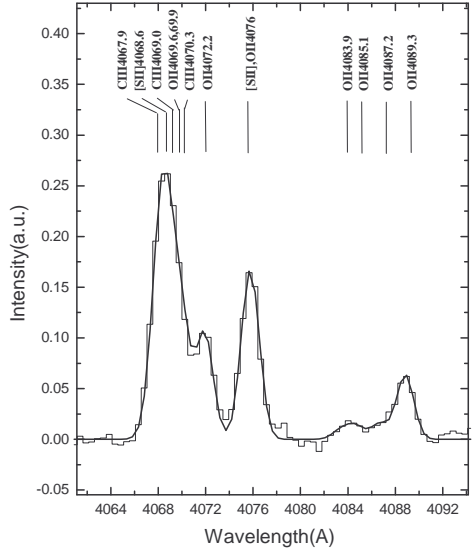, width=9 cm, clip=}
\caption{High resolution spectrum (1.5\,{\AA} FWHM) of NGC\,5882: the blend of
O~{\sc ii} quartet V\,10, C~{\sc iii} triplet V\,16 and {\fsii} doublet is well
fitted by multiple Gaussians (\emph{thick line}); also present are O~{\sc ii}
4f--3d transitions of which $\lambda$4089.3 ($J$~=~11/2--9/2) is the strongest
predicted line. See text for more details.}
\end{center}
\end{figure}

Table~3 presents the observed optical relative line intensities for the 12
Galactic PNe in our sample, while Table~4 presents the observed optical and
ultraviolet relative line intensities for the 3 Magellanic Cloud PNe. Table~5
presents the ultraviolet nebular line intensities for the 10 Galactic PNe in
the sample which were observed by the \emph{IUE}. Typical optical line flux
errors are estimated to be less than 5\,per cent for lines with observed fluxes
\emph{F}($\lambda$) $\geq$ 0.2 [in units of \emph{F}(H$\beta$)~=~100], 10\,per
cent for those with 0.1 $\leq$ \emph{F}($\lambda$) $\leq$ 0.2, 20\,per cent for
0.05 $\leq$ \emph{F}($\lambda$) $\leq$ 0.1, and 30\,per cent or more for lines
with \emph{F}($\lambda$) $\leq$ 0.05. For a given flux range, the errors for
optical lines shortwards of 3400\,{\AA} are slightly larger, due to the
decreasing CCD efficiency towards the end of the spectral coverage and the
effects of ozone absorption.


\setcounter{table}{2}
\begin{table}
\begin{minipage}{75mm}
\centering \caption{Observed and dereddened relative line fluxes
[$F$($\lambda$) and $I$($\lambda$), respectively], on a scale where
H$\beta~=~100$.}
\begin{tabular}{l@{\hspace{2.8mm}}l@{\hspace{2.8mm}}l@{\hspace{2.8mm}}c@{\hspace{2.8mm}}c@{\hspace{2.8mm}}c@{\hspace{2.8mm}}}
\noalign{\hrule} \noalign{\vskip3pt}
\multicolumn{6}{c}{NGC\,2022}\\
\multicolumn{6}{c}{\emph{c}(H$\beta$)$^\mathrm{{Ba}}$ = 0.38}\\
$\lambda_{\rm obs}$&$F(\lambda)$&$I(\lambda)$&ID&$\lambda_{\rm 0}$&Mult  \\
\noalign{\vskip3pt} \noalign{\hrule} \noalign{\vskip3pt}
             &       *&        * &[S III]  &3721.63&F2     \\%
             &       *&        * &H 14     &3721.94&H14    \\%
         &1.16($+$1)&1.45($+$1)  &[O II]   &3726.03&F1     \\
             &     *&          * &[O II]   &3728.82&F1     \\%
             &     *&          * &H 13     &3734.37&H13    \\
\end{tabular}
\end{minipage}
\end{table}

\setcounter{table}{2}
\begin{table}
\begin{minipage}{75mm}
\centering \caption{{\it --continued}}
\begin{tabular}{l@{\hspace{2.8mm}}l@{\hspace{2.8mm}}l@{\hspace{2.8mm}}c@{\hspace{2.8mm}}c@{\hspace{2.8mm}}c@{\hspace{2.8mm}}}
\noalign{\hrule} \noalign{\vskip3pt}
\multicolumn{6}{c}{NGC\,2022}\\
$\lambda_{\rm obs}$&$F(\lambda)$&$I(\lambda)$&ID&$\lambda_{\rm 0}$&Mult  \\
\noalign{\vskip3pt} \noalign{\hrule} \noalign{\vskip3pt}
 4621.81 &3.08($-$2)&3.24($-$2)&N II     &4621.81&V 5    \\%
 4626.09 &4.12($-$2)&4.33($-$2)&[Ar V]?  &4624.54&       \\
 4631.50 &3.18($-$1)&3.34($-$1)&O IV     &4631.89&V1     \\%
 4634.78 &5.34($-$1)&5.60($-$1)&N III    &4634.14&V2     \\%
 4639.50 &1.00($-$1)&1.05($-$1)&O II     &4638.86&V1     \\%
 4641.28 &8.87($-$1)&9.29($-$1)&N III    &4640.64&V2     \\
\end{tabular}
\end{minipage}
\end{table}

\setcounter{table}{2}
\begin{table}
\begin{minipage}{75mm}
\centering \caption{{\it --continued}}
\begin{tabular}{l@{\hspace{2.8mm}}l@{\hspace{2.8mm}}l@{\hspace{2.8mm}}c@{\hspace{2.8mm}}c@{\hspace{2.8mm}}c@{\hspace{2.8mm}}}
\noalign{\hrule} \noalign{\vskip3pt}
\multicolumn{6}{c}{NGC\,2022}\\
$\lambda_{\rm obs}$&$F(\lambda)$&$I(\lambda)$&ID&$\lambda_{\rm 0}$&Mult  \\
\noalign{\vskip3pt} \noalign{\hrule} \noalign{\vskip3pt}
 6862.86 &2.91($-$1)&2.13($-$1)&?        &       &       \\%
 6892.24 &1.06      &7.76($-$1)&He II    &6890.88&       \\%
 7007.02 &5.06      &3.64      &[Ar V]   &7005.67&       \\%
 7066.61 &6.42($-$1)&4.59($-$1)&He I     &7065.25&V10    \\%
 7136.82 &1.33($+$1)&9.51      &[Ar III] &7135.80&F1     \\%
 7171.65 &7.22($-$1)&5.11($-$1)&[Ar IV]  &7170.62&F2     \\
\end{tabular}
\end{minipage}
\end{table}

\setcounter{table}{2}
\begin{table}
\begin{minipage}{75mm}
\centering \caption{{\it --continued}}
\begin{tabular}{l@{\hspace{2.8mm}}l@{\hspace{2.8mm}}l@{\hspace{2.8mm}}c@{\hspace{2.8mm}}c@{\hspace{2.8mm}}c@{\hspace{2.8mm}}}
\noalign{\hrule} \noalign{\vskip3pt}
\multicolumn{6}{c}{NGC\,2440}\\
$\lambda_{\rm obs}$&$F(\lambda)$&$I(\lambda)$&ID&$\lambda_{\rm 0}$&Mult  \\
\noalign{\vskip3pt} \noalign{\hrule} \noalign{\vskip3pt}
 4621.31 &5.21($-$2)&5.54($-$2)&N II     &4621.39&V5   \\
 4626.41 &8.99($-$2)&9.55($-$2)&?        &       &     \\%
 4632.99 &2.82($-$1)&3.00($-$1)&O IV     &4631.89&V1   \\%
 4635.17 &2.07      &2.19      &N III    &4634.14&V2   \\%
 4639.89 &2.66($-$1)&2.84($-$1)&O II     &4638.86&V1   \\%
 4641.67 &3.79      &4.01      &N III    &4640.64&V2   \\
\end{tabular}
\end{minipage}
\end{table}

\setcounter{table}{2}
\begin{table}
\begin{minipage}{75mm}
\centering \caption{{\it --continued}}
\begin{tabular}{l@{\hspace{2.8mm}}l@{\hspace{2.8mm}}l@{\hspace{2.8mm}}c@{\hspace{2.8mm}}c@{\hspace{2.8mm}}c@{\hspace{2.8mm}}}
\noalign{\hrule} \noalign{\vskip3pt}
\multicolumn{6}{c}{NGC\,3132}\\
$\lambda_{\rm obs}$&$F(\lambda)$&$I(\lambda)$&ID&$\lambda_{\rm 0}$&Mult  \\
\noalign{\vskip3pt} \noalign{\hrule} \noalign{\vskip3pt}
 4120.42 &2.50($-$1)&2.71($-$1)&He I     &4120.84&V16    \\
       * &       * &       * &O II     &4121.46&V19    \\
 4143.38 &3.49($-$1)&3.78($-$1)&He I     &4143.76&V53    \\
 4153.25 &7.27($-$2)&7.86($-$2)&O II     &4153.30&V19    \\
 4168.53 &7.06($-$2)&7.63($-$2)&O II     &4169.22&V19    \\
 4266.77 &6.52($-$1)&6.97($-$1)&C II     &4267.15&V6     \\
\end{tabular}
\end{minipage}
\end{table}

\setcounter{table}{2}
\begin{table}
\begin{minipage}{75mm}
\centering \caption{{\it --continued}}
\begin{tabular}{l@{\hspace{2.8mm}}l@{\hspace{2.8mm}}l@{\hspace{2.8mm}}c@{\hspace{2.8mm}}c@{\hspace{2.8mm}}c@{\hspace{2.8mm}}}
\noalign{\hrule} \noalign{\vskip3pt}
\multicolumn{6}{c}{NGC\,3132}\\
$\lambda_{\rm obs}$&$F(\lambda)$&$I(\lambda)$&ID&$\lambda_{\rm 0}$&Mult  \\
\noalign{\vskip3pt} \noalign{\hrule} \noalign{\vskip3pt}
 5576.60 &7.37($-$1)&6.83($-$1)&[O I]    &5577.34&F3     \\
 5676.01 &2.70($-$2)&2.49($-$2)&N II    & 5676.02&V3     \\
 5678.50 &8.68($-$2)&7.99($-$2)&N II    & 5679.56&V3     \\
 5754.04 &8.91      &8.15      &[N II]   &5754.60&F3     \\
 5875.09 &1.80($+$1)&1.63($+$1)&He I     &5875.66&V11    \\
 6300.10 &4.21($+$1)&3.70($+$1)&[O I]    &6300.34&F1     \\
\end{tabular}
\end{minipage}
\end{table}

\setcounter{table}{2}
\begin{table}
\begin{minipage}{75mm}
\centering \caption{{\it --continued}}
\begin{tabular}{l@{\hspace{2.8mm}}l@{\hspace{2.8mm}}l@{\hspace{2.8mm}}c@{\hspace{2.8mm}}c@{\hspace{2.8mm}}c@{\hspace{2.8mm}}}
\noalign{\hrule} \noalign{\vskip3pt}
\multicolumn{6}{c}{NGC\,3242}\\
$\lambda_{\rm obs}$&$F(\lambda)$&$I(\lambda)$&ID&$\lambda_{\rm 0}$&Mult  \\
\noalign{\vskip3pt} \noalign{\hrule} \noalign{\vskip3pt}
 4101.52 &2.49($+$1)&2.59($+$1)&H 6      &4101.74&H6    \\%
 4115.52 &1.89($-$2)&1.97($-$2)&?        &       &      \\%
 4118.98 &6.50($-$2)&6.77($-$2)&O II     &4119.22&V20   \\%
 4120.60 &1.88($-$1)&1.95($-$1)&He I     &4120.84&V16   \\%
 4121.22 &2.71($-$2)&2.83($-$2)&O II     &4121.46&V19   \\%
 4128.49 &4.96($-$2)&5.17($-$2)&O II?    &4129.32&V19   \\
\end{tabular}
\end{minipage}
\end{table}

\setcounter{table}{2}
\begin{table}
\begin{minipage}{75mm}
\centering \caption{{\it --continued}}
\begin{tabular}{l@{\hspace{2.8mm}}l@{\hspace{2.8mm}}l@{\hspace{2.8mm}}c@{\hspace{2.8mm}}c@{\hspace{2.8mm}}c@{\hspace{2.8mm}}}
\noalign{\hrule} \noalign{\vskip3pt}
\multicolumn{6}{c}{NGC\,3242}\\
$\lambda_{\rm obs}$&$F(\lambda)$&$I(\lambda)$&ID&$\lambda_{\rm 0}$&Mult  \\
\noalign{\vskip3pt} \noalign{\hrule} \noalign{\vskip3pt}
 4442.12 &4.68($-$3)&4.80($-$3)&N II     &4442.02&V55a  \\%
 4448.11 &1.18($-$2)&1.21($-$2)&O II     &4448.19&V35   \\%
 4453.50 &1.74($-$2)&1.79($-$2)&O II     &4452.37&V5    \\%
 4458.18 &1.78($-$2)&1.82($-$2)&Ne II    &4457.24&V61d  \\%
 4465.79 &2.54($-$2)&2.60($-$2)&O II     &4466.42&V86b  \\%
 4471.29 &4.03      &4.12      &He I     &4471.50&V14   \\
\end{tabular}
\end{minipage}
\end{table}

\setcounter{table}{2}
\begin{table}
\begin{minipage}{75mm}
\centering \caption{{\it --continued}}
\begin{tabular}{l@{\hspace{2.8mm}}l@{\hspace{2.8mm}}l@{\hspace{2.8mm}}c@{\hspace{2.8mm}}c@{\hspace{2.8mm}}c@{\hspace{2.8mm}}}
\noalign{\hrule} \noalign{\vskip3pt}
\multicolumn{6}{c}{NGC\,3242}\\
$\lambda_{\rm obs}$&$F(\lambda)$&$I(\lambda)$&ID&$\lambda_{\rm 0}$&Mult  \\
\noalign{\vskip3pt} \noalign{\hrule} \noalign{\vskip3pt}
 5345.20 &7.80($-$2)&7.59($-$2)&[Kr IV]  &5345.90 &     \\
 5411.06 &1.97      &1.91      &He II    &5411.52& 4.7  \\%
 5517.36 &3.29($-$1)&3.17($-$1)&[Cl III] &5517.66&F1    \\%
 5537.29 &2.88($-$1)&2.77($-$1)&[Cl III] &5537.60&F1    \\%
 5576.67 &3.70($-$2)&3.56($-$2)&[O I]    &5577.34&F3    \\%
 5591.52 &5.27($-$2)&5.07($-$2)&O III    &5592.37&V5    \\
\end{tabular}
\end{minipage}
\end{table}

\setcounter{table}{2}
\begin{table}
\begin{minipage}{75mm}
\centering \caption{{\it --continued}}
\begin{tabular}{l@{\hspace{2.8mm}}l@{\hspace{2.8mm}}l@{\hspace{2.8mm}}c@{\hspace{2.8mm}}c@{\hspace{2.8mm}}c@{\hspace{2.8mm}}}
\noalign{\hrule} \noalign{\vskip3pt}
\multicolumn{6}{c}{NGC\,3918}\\
$\lambda_{\rm obs}$&$F(\lambda)$&$I(\lambda)$&ID&$\lambda_{\rm 0}$&Mult  \\
\noalign{\vskip3pt} \noalign{\hrule} \noalign{\vskip3pt}
 3834.62 &6.09      &7.73      &H 9      &3835.39&H9    \\%
 3867.98 &1.07($+$2)&1.35($+$2)&[Ne III] &3868.75&F1    \\%
 3888.09 &1.57($+$1)&1.98($+$1)&He I     &3888.65&V2    \\%
 3922.02 &2.50($-$1)&3.11($-$1)&He II    &3923.48&4.15  \\%
 3967.20 &4.65($+$1)&5.75($+$1)&[Ne III] &3967.46&F1    \\%
 4008.79 &1.10($-$1)&1.34($-$1)&He I     &4009.26&V55   \\
\end{tabular}
\end{minipage}
\end{table}

\setcounter{table}{2}
\begin{table}
\begin{minipage}{75mm}
\centering \caption{{\it --continued}}
\begin{tabular}{l@{\hspace{2.8mm}}l@{\hspace{2.8mm}}l@{\hspace{2.8mm}}c@{\hspace{2.8mm}}c@{\hspace{2.8mm}}c@{\hspace{2.8mm}}}
\noalign{\hrule} \noalign{\vskip3pt}
\multicolumn{6}{c}{NGC\,3918}\\
$\lambda_{\rm obs}$&$F(\lambda)$&$I(\lambda)$&ID&$\lambda_{\rm 0}$&Mult  \\
\noalign{\vskip3pt} \noalign{\hrule} \noalign{\vskip3pt}
 4448.19 &1.59($-$2)&1.76($-$2)&O II     &4448.19&V35   \\%
 4452.37 &5.90($-$2)&6.53($-$2)&O II     &4452.37&V5    \\%
 4458.10 &3.40($-$2)&3.75($-$2)&Ne II    &4457.05&V61d  \\
 4465.13 &2.20($-$2)&2.42($-$2)&O II     &4466.42&V86b  \\
 4470.93 &3.53      &3.88      &He I     &4471.50&V14   \\%
 4480.62 &1.10($-$2)&1.20($-$2)&Mg II    &4481.21&V4    \\
\end{tabular}
\end{minipage}
\end{table}

\setcounter{table}{2}
\begin{table}
\begin{minipage}{75mm}
\centering \caption{{\it --continued}}
\begin{tabular}{l@{\hspace{2.8mm}}l@{\hspace{2.8mm}}l@{\hspace{2.8mm}}c@{\hspace{2.8mm}}c@{\hspace{2.8mm}}c@{\hspace{2.8mm}}}
\noalign{\hrule} \noalign{\vskip3pt}
\multicolumn{6}{c}{NGC\,3918}\\
$\lambda_{\rm obs}$&$F(\lambda)$&$I(\lambda)$&ID&$\lambda_{\rm 0}$&Mult  \\
\noalign{\vskip3pt} \noalign{\hrule} \noalign{\vskip3pt}
 5591.58 &7.60($-$2)&6.42($-$2)& O III   &5592.37&V5    \\%
 5753.69 &2.11      &1.73      &[N II]   &5754.60&F3    \\%
 5800.81 &1.20($-$1)&9.78($-$2)&C IV     &5801.51&V1    \\%
 5874.81 &1.34($+$1)&1.08($+$1)&He I     &5875.66&V11   \\%
 6086.35 &1.40($-$1)&1.09($-$1)&[Fe VII] &6086.90&      \\
 6101.17 &4.50($-$1)&3.49($-$1)&[K IV]   &6101.83&F1    \\
\end{tabular}
\end{minipage}
\end{table}

\setcounter{table}{2}
\begin{table}
\begin{minipage}{75mm}
\centering \caption{{\it --continued}}
\begin{tabular}{l@{\hspace{2.8mm}}l@{\hspace{2.8mm}}l@{\hspace{2.8mm}}c@{\hspace{2.8mm}}c@{\hspace{2.8mm}}c@{\hspace{2.8mm}}}
\noalign{\hrule} \noalign{\vskip3pt}
\multicolumn{6}{c}{NGC\,5315}\\
$\lambda_{\rm obs}$&$F(\lambda)$&$I(\lambda)$&ID&$\lambda_{\rm 0}$&Mult  \\
\noalign{\vskip3pt} \noalign{\hrule} \noalign{\vskip3pt}
         &        * &      *      &O II     &4120.54&V20 \\%
 4121.32 &3.90($-$2)&5.20($-$2)&He I     &4120.84&V16 \\%
         &      *   &      *      &O II     &4121.46&V19 \\%
 4131.68 &5.50($-$2)&7.29($-$2)&O II     &4132.80&V19 \\%
 4142.89 &3.80($-$1)&5.01($-$1)&He I     &4143.76&V53 \\%
 4152.27 &6.92($-$2)&9.10($-$2)&O II     &4153.30&V19 \\
\end{tabular}
\end{minipage}
\end{table}

\setcounter{table}{2}
\begin{table}
\begin{minipage}{75mm}
\centering \caption{{\it --continued}}
\begin{tabular}{l@{\hspace{2.8mm}}l@{\hspace{2.8mm}}l@{\hspace{2.8mm}}c@{\hspace{2.8mm}}c@{\hspace{2.8mm}}c@{\hspace{2.8mm}}}
\noalign{\hrule} \noalign{\vskip3pt}
\multicolumn{6}{c}{NGC\,5315}\\
$\lambda_{\rm obs}$&$F(\lambda)$&$I(\lambda)$&ID&$\lambda_{\rm 0}$&Mult  \\
\noalign{\vskip3pt} \noalign{\hrule} \noalign{\vskip3pt}
 4698.92 &2.34($-$2)&2.49($-$2)&O II     &4699.22&V25 \\%
 4700.66 &6.72($-$2)&7.15($-$2)&[Fe III] &4701.62&F3  \\%
 4712.20 &7.90($-$1)&8.38($-$1)&[Ar IV]  &4711.37&F1  \\
 4732.86 &3.70($-$2)&3.89($-$2)&?        &       &    \\
 4739.46 &1.80($-$1)&1.88($-$1)&[Ar IV]  &4740.17&F1  \\%
 4753.58 &4.50($-$3)&4.69($-$3)&[Fe III] &4754.83&    \\
\end{tabular}
\end{minipage}
\end{table}

\setcounter{table}{2}
\begin{table}
\begin{minipage}{75mm}
\centering \caption{{\it --continued}}
\begin{tabular}{l@{\hspace{2.8mm}}l@{\hspace{2.8mm}}l@{\hspace{2.8mm}}c@{\hspace{2.8mm}}c@{\hspace{2.8mm}}c@{\hspace{2.8mm}}}
\noalign{\hrule} \noalign{\vskip3pt}
\multicolumn{6}{c}{NGC\,5882}\\
$\lambda_{\rm obs}$&$F(\lambda)$&$I(\lambda)$&ID&$\lambda_{\rm 0}$&Mult  \\
\noalign{\vskip3pt} \noalign{\hrule} \noalign{\vskip3pt}
 3722.16 &2.15      &2.94      &H 14     &3721.94&H14  \\%
 3726.41 &7.49      &1.02($+$1)&[O II]   &3726.03&F1   \\%
 3729.20 &3.65      &4.98      &[O II]   &3728.82&F1   \\%
 3734.75 &1.68      &2.29      &H 13     &3734.37&H13  \\%
 3750.50 &2.24      &3.04      &H 12     &3750.15&H12  \\%
 3755.09 &2.89($-$1)&3.92($-$1)&O III    &3754.69&V2   \\
\end{tabular}
\end{minipage}
\end{table}

\setcounter{table}{2}
\begin{table}
\begin{minipage}{75mm}
\centering \caption{{\it --continued}}
\begin{tabular}{l@{\hspace{2.8mm}}l@{\hspace{2.8mm}}l@{\hspace{2.8mm}}c@{\hspace{2.8mm}}c@{\hspace{2.8mm}}c@{\hspace{2.8mm}}}
\noalign{\hrule} \noalign{\vskip3pt}
\multicolumn{6}{c}{NGC\,5882}\\
$\lambda_{\rm obs}$&$F(\lambda)$&$I(\lambda)$&ID&$\lambda_{\rm 0}$&Mult  \\
\noalign{\vskip3pt} \noalign{\hrule} \noalign{\vskip3pt}
 4277.48 &2.22($-$2)&2.65($-$2)&O II     &4277.43&V67c \\
 4291.14 &1.87($-$2)&2.21($-$2)&O II     &4291.25&V55  \\%
 4292.10 &6.59($-$3)&7.79($-$3)&O II     &4292.21&V18c \\%
 4294.67 &1.32($-$2)&1.56($-$2)&O II     &4294.78&V53b \\%
 4303.51 &4.86($-$2)&5.73($-$2)&O II     &4303.61&V65a \\
 4317.11 &5.43($-$2)&6.38($-$2)&O II     &4317.14&V2   \\
\end{tabular}
\end{minipage}
\end{table}

\setcounter{table}{2}
\begin{table}
\begin{minipage}{75mm}
\centering \caption{{\it --continued}}
\begin{tabular}{l@{\hspace{2.8mm}}l@{\hspace{2.8mm}}l@{\hspace{2.8mm}}c@{\hspace{2.8mm}}c@{\hspace{2.8mm}}c@{\hspace{2.8mm}}}
\noalign{\hrule} \noalign{\vskip3pt}
\multicolumn{6}{c}{NGC\,5882}\\
$\lambda_{\rm obs}$&$F(\lambda)$&$I(\lambda)$&ID&$\lambda_{\rm 0}$&Mult  \\
\noalign{\vskip3pt} \noalign{\hrule} \noalign{\vskip3pt}
 4802.00 &9.80($-$3)&9.98($-$3)&C II     &4802.23&     \\%
 4803.15 &1.80($-$2)&1.83($-$2)&N II    & 4803.29&V20  \\
 4860.80 &1.00($+$2)&1.00($+$2)&H 4      &4861.33&H4   \\%
 4881.03 &6.38($-$2)&6.34($-$2)&[Fe III] &4881.11&F2   \\%
 4891.60 &8.74($-$3)&8.66($-$3)&O II     &4890.86&V28  \\%
 4899.28 &8.78($-$3)&8.68($-$3)&Fe IV?   &       &     \\
\end{tabular}
\end{minipage}
\end{table}

\clearpage

\setcounter{table}{2}
\begin{table}
\begin{minipage}{75mm}
\centering \caption{{\it --continued}}
\begin{tabular}{l@{\hspace{2.8mm}}l@{\hspace{2.8mm}}l@{\hspace{2.8mm}}c@{\hspace{2.8mm}}c@{\hspace{2.8mm}}c@{\hspace{2.8mm}}}
\noalign{\hrule} \noalign{\vskip3pt}
\multicolumn{6}{c}{NGC\,6302}\\
\multicolumn{6}{c}{\emph{c}(H$\beta$)$^\mathrm{{Ba}}$ = 1.44} \\
$\lambda_{\rm obs}$&$F(\lambda)$&$I(\lambda)$&ID&$\lambda_{\rm 0}$&Mult  \\
\noalign{\vskip3pt} \noalign{\hrule} \noalign{\vskip3pt}

 3118.22 &1.16      &5.24      &         &      *&    \\%
 3120.54 &1.31      &5.86      &O III    &3121.71&V12 \\%
 3131.16 &1.66      &7.33      &O III    &3132.79&V12 \\%
 3184.38 &1.21($-$1)&4.96($-$1)&Si III   &3185.12&V8  \\%
 3187.00 &8.32($-$1)&3.38      &He I     &3187.74&V3  \\
\end{tabular}
\end{minipage}
\end{table}

\setcounter{table}{2}
\begin{table}
\begin{minipage}{75mm}
\centering \caption{{\it --continued}}
\begin{tabular}{l@{\hspace{2.8mm}}l@{\hspace{2.8mm}}l@{\hspace{2.8mm}}c@{\hspace{2.8mm}}c@{\hspace{2.8mm}}c@{\hspace{2.8mm}}}
\noalign{\hrule} \noalign{\vskip3pt}
\multicolumn{6}{c}{NGC\,6302}\\
$\lambda_{\rm obs}$&$F(\lambda)$&$I(\lambda)$&ID&$\lambda_{\rm 0}$&Mult  \\
\noalign{\vskip3pt} \noalign{\hrule} \noalign{\vskip3pt}
 3857.65 &1.52($-$1)&3.28($-$1)&He II    &3858.07&4.17\\%
 3888.47 &8.57      &1.81($+$1)&He I     &3888.65&V2  \\%
 3895.42 &1.01($-$2)&2.13($-$2)&?        &       &    \\%
 3923.18 &2.15($-$1)&4.45($-$1)&He II    &3923.48&4.15\\%
 3926.23 &5.07($-$1)&1.04      &He I     &3926.54&V58 \\%
 3967.11 &2.76($+$1)&5.56($+$1)&[Ne III] &3967.46&F1  \\
\end{tabular}
\end{minipage}
\end{table}

\setcounter{table}{2}
\begin{table}
\begin{minipage}{75mm}
\centering \caption{{\it --continued}}
\begin{tabular}{l@{\hspace{2.8mm}}l@{\hspace{2.8mm}}l@{\hspace{2.8mm}}c@{\hspace{2.8mm}}c@{\hspace{2.8mm}}c@{\hspace{2.8mm}}}
\noalign{\hrule} \noalign{\vskip3pt}
\multicolumn{6}{c}{NGC\,6302}\\
$\lambda_{\rm obs}$&$F(\lambda)$&$I(\lambda)$&ID&$\lambda_{\rm 0}$&Mult  \\
\noalign{\vskip3pt} \noalign{\hrule} \noalign{\vskip3pt}

 4880.73 &9.12($-$2)&8.97($-$2)&[Fe III] &4881.11&F2     \\%
 4893.40 &5.44($-$2)&5.30($-$2)&?        &       &       \\%
 4921.75 &1.52      &1.45      &He I     &4921.93&V48    \\%
 4930.63 &5.59($-$1)&5.28($-$1)&O III    &4931.80&F1     \\%
 4934.69 &5.91($-$2)&5.56($-$2)&?        &       &       \\%
 4944.42 &2.40($-$1)&2.25($-$1)&O II     &4943.00&V33    \\
\end{tabular}
\end{minipage}
\end{table}

\setcounter{table}{2}
\begin{table}
\begin{minipage}{75mm}
\centering \caption{{\it --continued}}
\begin{tabular}{l@{\hspace{2.8mm}}l@{\hspace{2.8mm}}l@{\hspace{2.8mm}}c@{\hspace{2.8mm}}c@{\hspace{2.8mm}}c@{\hspace{2.8mm}}}
\noalign{\hrule} \noalign{\vskip3pt}
\multicolumn{6}{c}{NGC\,6302}\\
$\lambda_{\rm obs}$&$F(\lambda)$&$I(\lambda)$&ID&$\lambda_{\rm 0}$&Mult  \\
\noalign{\vskip3pt} \noalign{\hrule} \noalign{\vskip3pt}

 7170.06 &4.73      &1.25      &[Ar IV]  &7170.62&F2     \\%
 7176.94 &2.81      &7.44($-$1)&He II    &7177.50&5.11   \\%
 7236.17 &3.40      &8.77($-$1)&C II     &7236.42&V3     \\%
         &*         &*         &[Ar IV]  &7237.40&F2     \\%
 7262.51 &3.79      &9.68($-$1)&[Ar IV]  &7262.76&F2     \\%
 7281.09 &3.52      &8.93($-$1)&He I     &7281.35&V45    \\
\end{tabular}
\end{minipage}
\end{table}

\setcounter{table}{2}
\begin{table}
\begin{minipage}{75mm}
\centering \caption{{\it --continued}}
\begin{tabular}{l@{\hspace{2.8mm}}l@{\hspace{2.8mm}}l@{\hspace{2.8mm}}c@{\hspace{2.8mm}}c@{\hspace{2.8mm}}c@{\hspace{2.8mm}}}
\noalign{\hrule} \noalign{\vskip3pt}
\multicolumn{6}{c}{NGC\,6818}\\
$\lambda_{\rm obs}$&$F(\lambda)$&$I(\lambda)$&ID&$\lambda_{\rm 0}$&Mult  \\
\noalign{\vskip3pt} \noalign{\hrule} \noalign{\vskip3pt}
 3923.29 &3.00($-$1)&3.81($-$1)&He II    &3923.48&4.15  \\%
 3967.23 &4.78($+$1)&6.01($+$1)&[Ne III] &3967.46&F1    \\%
 4008.82 &7.40($-$2)&9.21($-$2)&He I     &4009.26&V55   \\%
 4025.71 &1.52      &1.88      &He I     &4026.21&V18   \\%
 4059.90 &3.60($-$2)&4.43($-$2)&[F IV]   &4060.23&      \\%
 4067.64 &1.93($-$1)&2.38($-$1)&C III    &4067.94&V16   \\
\end{tabular}
\end{minipage}
\end{table}

\setcounter{table}{2}
\begin{table}
\begin{minipage}{75mm}
\centering \caption{{\it --continued}}
\begin{tabular}{l@{\hspace{2.8mm}}l@{\hspace{2.8mm}}l@{\hspace{2.8mm}}c@{\hspace{2.8mm}}c@{\hspace{2.8mm}}c@{\hspace{2.8mm}}}
\noalign{\hrule} \noalign{\vskip3pt}
\multicolumn{6}{c}{NGC\,6818}\\
$\lambda_{\rm obs}$&$F(\lambda)$&$I(\lambda)$&ID&$\lambda_{\rm 0}$&Mult  \\
\noalign{\vskip3pt} \noalign{\hrule} \noalign{\vskip3pt}
 4713.85 &7.90($-$1)&8.22($-$1)&He I     &4713.17&V12   \\%
 4724.04 &6.48($-$1)&6.72($-$1)&[Ne IV]  &4724.15&F1    \\%
 4725.51 &5.19($-$1)&5.38($-$1)&[Ne IV]  &4725.62&F1    \\%
 4740.12 &7.10      &7.33      &[Ar IV]  &4740.17&F1    \\%
 4796.03 &4.10($-$2)&4.17($-$2)&?        &       &      \\
 4831.07 &2.70($-$2)&2.72($-$2)&?        &       &      \\
\end{tabular}
\end{minipage}
\end{table}

\setcounter{table}{2}
\begin{table}
\begin{minipage}{75mm}
\centering \caption{{\it --continued}}
\begin{tabular}{l@{\hspace{2.8mm}}l@{\hspace{2.8mm}}l@{\hspace{2.8mm}}c@{\hspace{2.8mm}}c@{\hspace{2.8mm}}c@{\hspace{2.8mm}}}
\noalign{\hrule} \noalign{\vskip3pt}
\multicolumn{6}{c}{IC\,4191 (fixed-slit)}\\
\multicolumn{6}{c}{\emph{c}(H$\beta$)$^\mathrm{{Ba}}$ = 0.61} \\
$\lambda_{\rm obs}$&$F(\lambda)$&$I(\lambda)$&ID&$\lambda_{\rm 0}$&Mult  \\
\noalign{\vskip3pt} \noalign{\hrule} \noalign{\vskip3pt}
 3704.18 &7.17($-$1)&1.03      &He I     &3705.02&     \\%
 3711.95 &6.00($-$1)&8.63($-$1)&H 15     &3711.97&     \\%
       * &       *&      * &H 14     &3721.94&H14  \\%
       * &       *&      * &[S III]  &3721.63&F2   \\%
 3726.19 &3.24($+$1)&4.65($+$1)&[O II]   &3726.03&F1   \\
\end{tabular}
\end{minipage}
\end{table}

\setcounter{table}{2}
\begin{table}
\begin{minipage}{75mm}
\centering \caption{{\it --continued}}
\begin{tabular}{l@{\hspace{2.8mm}}l@{\hspace{2.8mm}}l@{\hspace{2.8mm}}c@{\hspace{2.8mm}}c@{\hspace{2.8mm}}c@{\hspace{2.8mm}}}
\noalign{\hrule} \noalign{\vskip3pt}
\multicolumn{6}{c}{IC\,4191 (fixed-slit)}\\
$\lambda_{\rm obs}$&$F(\lambda)$&$I(\lambda)$&ID&$\lambda_{\rm 0}$&Mult  \\
\noalign{\vskip3pt} \noalign{\hrule} \noalign{\vskip3pt}

 4590.82 &7.32($-$2)&8.04($-$2)&O II     &4590.97&     \\%
 4596.03 &5.63($-$2)&6.17($-$2)&O II     &4596.18&V15  \\%
 4609.75 &5.52($-$2)&6.03($-$2)&O II     &4609.44&V92a \\%
         &       *&      *     &O II     &4610.20&V92c \\%
 4631.93 &7.53($-$2)&8.15($-$2)&N II     &4630.54&V5   \\%
 4634.03 &9.33($-$1)&1.01      &N III    &4634.14&V2   \\
\end{tabular}
\end{minipage}
\end{table}

\setcounter{table}{2}
\begin{table}
\begin{minipage}{75mm}
\centering \caption{{\it --continued}}
\begin{tabular}{l@{\hspace{2.8mm}}l@{\hspace{2.8mm}}l@{\hspace{2.8mm}}c@{\hspace{2.8mm}}c@{\hspace{2.8mm}}c@{\hspace{2.8mm}}}
\noalign{\hrule} \noalign{\vskip3pt}
\multicolumn{6}{c}{IC\,4191 (fixed-slit)}\\
$\lambda_{\rm obs}$&$F(\lambda)$&$I(\lambda)$&ID&$\lambda_{\rm 0}$&Mult  \\
\noalign{\vskip3pt} \noalign{\hrule} \noalign{\vskip3pt}
 6347.23 &2.76($-$1)&1.84($-$1)&Si II    &6347.10&V2   \\%
 6363.91 &4.07      &2.70      &[O I]    &6363.78&F1   \\%
 6371.51 &2.23($-$1)&1.48($-$1)&Si II    &6371.38&V2   \\%
 6393.98 &1.17($-$1)&7.78($-$2)&[Mn V]   &6393.60&     \\%
 6401.95 &9.58($-$2)&6.31($-$2)&?        &       &     \\%
 6406.68 &1.26($-$1)&8.33($-$2)&He II    &6406.30&5.15 \\
\end{tabular}
\end{minipage}
\end{table}

\setcounter{table}{2}
\begin{table}
\begin{minipage}{75mm}
\centering \caption{{\it --continued}}
\begin{tabular}{l@{\hspace{2.8mm}}l@{\hspace{2.8mm}}l@{\hspace{2.8mm}}c@{\hspace{2.8mm}}c@{\hspace{2.8mm}}c@{\hspace{2.8mm}}}
\noalign{\hrule} \noalign{\vskip3pt}
\multicolumn{6}{c}{IC\,4191 (scanning-slit)}\\
$\lambda_{\rm obs}$&$F(\lambda)$&$I(\lambda)$&ID&$\lambda_{\rm 0}$&Mult  \\
\noalign{\vskip3pt} \noalign{\hrule} \noalign{\vskip3pt}
 4120.32 &1.95($-$1)&2.48($-$1)&He I     &4120.84&V16  \\%
 4143.07 &3.17($-$1)&3.99($-$1)&He I     &4143.76&V53  \\%
 4184.56 &3.60($-$2)&4.48($-$2)&O II     &4185.45&V36  \\%
 4186.01 &5.72($-$2)&7.11($-$2)&C III    &4186.90&V18  \\%
 4188.90 &5.71($-$2)&7.09($-$2)&O II     &4189.79&V36  \\%
 4199.27 &2.23($-$1)&2.76($-$1)&He II    &4199.83& 4.11\\
\end{tabular}
\end{minipage}
\end{table}

\setcounter{table}{2}
\begin{table}
\begin{minipage}{75mm}
\centering \caption{{\it --continued}}
\begin{tabular}{l@{\hspace{2.8mm}}l@{\hspace{2.8mm}}l@{\hspace{2.8mm}}c@{\hspace{2.8mm}}c@{\hspace{2.8mm}}c@{\hspace{2.8mm}}}
\noalign{\hrule} \noalign{\vskip3pt}
\multicolumn{6}{c}{IC\,4191 (scanning-slit)}\\
$\lambda_{\rm obs}$&$F(\lambda)$&$I(\lambda)$&ID&$\lambda_{\rm 0}$&Mult  \\
\noalign{\vskip3pt} \noalign{\hrule} \noalign{\vskip3pt}
 5576.38 &1.82($-$1)&1.46($-$1)&[O I]    &5577.34&F3   \\%
 5666.55 &1.17($-$1)&9.21($-$2)&N II    & 5666.63&V3   \\
 5677.93 &5.08($-$2)&4.00($-$2)&N II     &5676.02&V3   \\%
 5679.41 &1.63($-$1)&1.28($-$1)&N II    & 5679.56&V3    \\
 5753.56 &3.01      &2.32      &[N II]   &5754.60&F3   \\%
 5874.60 &2.29($+$1)&1.72($+$1)&He I     &5875.66&V11  \\
\end{tabular}
\end{minipage}
\end{table}

\setcounter{table}{2}
\begin{table}
\begin{minipage}{75mm}
\centering \caption{{\it --continued}}
\begin{tabular}{l@{\hspace{2.8mm}}l@{\hspace{2.8mm}}l@{\hspace{2.8mm}}c@{\hspace{2.8mm}}c@{\hspace{2.8mm}}c@{\hspace{2.8mm}}}
\noalign{\hrule} \noalign{\vskip3pt}
\multicolumn{6}{c}{IC\,4406}\\
$\lambda_{\rm obs}$&$F(\lambda)$&$I(\lambda)$&ID&$\lambda_{\rm 0}$&Mult  \\
\noalign{\vskip3pt} \noalign{\hrule} \noalign{\vskip3pt}
 4199.56 &3.02($-$1)&3.55($-$1)&He II    &4199.83& 4.11\\%
 4266.82 &6.96($-$1)&8.05($-$1)&C II     &4267.15&V6   \\%
 4325.24 &1.07($-$1)&1.23($-$1)&O II     &4325.76&V2   \\%
 4340.09 &4.02($+$1)&4.57($+$1)&H 5      &4340.47&H5   \\%
 4362.83 &6.00      &6.79      &[O III]  &4363.21&F2   \\%
 4378.89 &1.52($-$1)&1.71($-$1)&N III    &4379.11&V18  \\
\end{tabular}
\end{minipage}
\end{table}

\setcounter{table}{2}
\begin{table}
\begin{minipage}{75mm}
\centering \caption{{\it --continued}}
\begin{tabular}{l@{\hspace{2.8mm}}l@{\hspace{2.8mm}}l@{\hspace{2.8mm}}c@{\hspace{2.8mm}}c@{\hspace{2.8mm}}c@{\hspace{2.8mm}}}
\noalign{\hrule} \noalign{\vskip3pt}
\multicolumn{6}{c}{My\,Cn\,18}\\
\multicolumn{6}{c}{\emph{c}(H$\beta$)$^\mathrm{{Ba}}$ = 1.24} \\
$\lambda_{\rm obs}$&$F(\lambda)$&$I(\lambda)$&ID&$\lambda_{\rm 0}$&Mult  \\
\noalign{\vskip3pt} \noalign{\hrule} \noalign{\vskip3pt}
         &      * & *      &H 14     &3721.94&H14   \\%
         &      * & *      &[S III]  &3721.63&F2    \\%
 3724.96 &3.05($+$1)&6.36($+$1)&[O II]   &3726.03&F1    \\
         &      * & *      &[O II]   &3728.82&F1    \\%
         &      * & *      &H 13     &3734.37&H13   \\
\end{tabular}
\end{minipage}
\end{table}

\setcounter{table}{2}
\begin{table}
\begin{minipage}{75mm}
\centering \caption{{\it --continued}}
\begin{tabular}{l@{\hspace{2.8mm}}l@{\hspace{2.8mm}}l@{\hspace{2.8mm}}c@{\hspace{2.8mm}}c@{\hspace{2.8mm}}c@{\hspace{2.8mm}}}
\noalign{\hrule} \noalign{\vskip3pt}
\multicolumn{6}{c}{My\,Cn\,18}\\
\multicolumn{6}{c}{\emph{c}(H$\beta$)$^\mathrm{{Ba}}$ = 1.24} \\
$\lambda_{\rm obs}$&$F(\lambda)$&$I(\lambda)$&ID&$\lambda_{\rm 0}$&Mult  \\
\noalign{\vskip3pt} \noalign{\hrule} \noalign{\vskip3pt}

 6547.09 &1.43($+$2)&5.79($+$1)&[N II]   &6548.10&F1    \\%
 6561.75 &5.73($+$2)&2.29($+$2)&H 3      &6562.77&H3    \\%
 6582.48 &4.42($+$2)&1.76($+$2)&[N II]   &6583.50&F1    \\%
 6677.03 &9.92      &3.79      &He I     &6678.16&V46   \\%
 6715.42 &7.58      &2.86      &[S II]   &6716.44&F2    \\%
 6729.80 &1.34($+$1)&5.04      &[S II]   &6730.82&F2    \\%
 7063.98 &1.54($+$1)&5.11      &He I     &7065.25&V10   \\%
 7134.51 &3.43($+$1)&1.10($+$1)&[Ar III] &7135.80&F1    \\%
 7176.65 &1.01      &3.23($-$1)&He II    &7177.50&5.11  \\%
 7230.02 &2.47($-$1)&7.71($-$2)&C II     &7231.32&V3    \\%
 7235.12 &6.20($-$1)&1.93($-$1)&C II     &7236.42&V3    \\%
 7279.85 &1.65      &5.08($-$1)&He I     &7281.35&V45   \\
 7318.18 &1.19($+$1)&3.61      &[O II]   &7318.92&F2    \\%
 7328.93 &1.00($+$1)&3.02      &[O II]   &7329.67&F2    \\%
\noalign{\vskip3pt} \noalign{\hrule} \noalign{\vskip3pt}
\end{tabular}
\end{minipage}
\end{table}

\setcounter{table}{3}
\begin{table}
\begin{minipage}{75mm}
\centering \caption{Observed and dereddened relative line fluxes
[$F$($\lambda$) and $I$($\lambda$), respectively], on a scale where
H$\beta~=~100$.}
\begin{tabular}{l@{\hspace{2.8mm}}l@{\hspace{2.8mm}}l@{\hspace{2.8mm}}c@{\hspace{2.8mm}}c@{\hspace{2.8mm}}c@{\hspace{2.8mm}}}
\noalign{\hrule} \noalign{\vskip3pt}
\multicolumn{6}{c}{SMC~N87}\\
\multicolumn{6}{c}{\emph{c}(H$\beta$)$^\mathrm{{Ba}}$ = 0.03 (Gal), 0.02 (SMC)} \\
$\lambda_{\rm obs}$&$F(\lambda)$&$I(\lambda)$&ID&$\lambda_{\rm 0}$&Mult  \\
\noalign{\vskip3pt} \noalign{\hrule} \noalign{\vskip3pt}
 1549    &49.7      &62.4      &C IV      &1549   &\\
 1663    &19.3      &23.8      &O III]    &1663   &\\
 1908    &451       &550       &C III]    &1908   &\\
 2426    &52.1      &62.4      &C II]     &2426   &\\
 3721.72 &1.26      &1.61      &H 14     &3721.94&H14  \\
\end{tabular}
\end{minipage}
\end{table}

\setcounter{table}{3}
\begin{table}
\begin{minipage}{75mm}
\centering \caption{{\it --continued}}
\begin{tabular}{l@{\hspace{2.8mm}}l@{\hspace{2.8mm}}l@{\hspace{2.8mm}}c@{\hspace{2.8mm}}c@{\hspace{2.8mm}}c@{\hspace{2.8mm}}}
\noalign{\hrule} \noalign{\vskip3pt}
\multicolumn{6}{c}{SMC~N87}\\
$\lambda_{\rm obs}$&$F(\lambda)$&$I(\lambda)$&ID&$\lambda_{\rm 0}$&Mult  \\
\noalign{\vskip3pt} \noalign{\hrule} \noalign{\vskip3pt}
 4804.79 &3.47($-$2)&3.50($-$2)&N II     &4802.29&V20  \\
 4861.30 &1.00($+$2)&1.00($+$2)&H 4      &4861.33&H4   \\
 4923.89 &1.30      &1.28      &He I     &4921.93&V48  \\
 4926.49 &6.55($-$2)&6.48($-$2)&O II     &4924.53&V28  \\
 4933.09 &5.84($-$2)&5.79($-$2)&[O III]  &4931.80&F1   \\
 4960.99 &1.93($+$2)&1.90($+$2)&[O III]  &4958.91&F1   \\
\end{tabular}
\end{minipage}
\end{table}

\setcounter{table}{3}
\begin{table}
\begin{minipage}{75mm}
\centering \caption{{\it --continued}}
\begin{tabular}{l@{\hspace{2.8mm}}l@{\hspace{2.8mm}}l@{\hspace{2.8mm}}c@{\hspace{2.8mm}}c@{\hspace{2.8mm}}c@{\hspace{2.8mm}}}
\noalign{\hrule} \noalign{\vskip3pt}
\multicolumn{6}{c}{LMC~N141}\\
$\lambda_{\rm obs}$&$F(\lambda)$&$I(\lambda)$&ID&$\lambda_{\rm 0}$&Mult  \\
\noalign{\vskip3pt} \noalign{\hrule} \noalign{\vskip3pt}
 4665.65 &5.00($-$2)&5.18($-$2)&O II     &4661.63&V1   \\
 4680.62 &4.50($-$2)&4.65($-$2)&O II     &4676.24&V1   \\
 4689.66 &3.03($-$1)&3.13($-$1)&He II    &4685.68& 3.4 \\
 4705.54 &5.63($-$2)&5.78($-$2)&[Fe III] &4701.62&F3   \\
 4715.34 &8.29($-$1)&8.51($-$1)&[Ar IV]  &4711.37&F1   \\
 4717.14 &8.59($-$1)&8.81($-$1)&He I     &4713.17&v12  \\
\end{tabular}
\end{minipage}
\end{table}

\setcounter{table}{3}
\begin{table}
\begin{minipage}{75mm}
\centering \caption{{\it --continued}}
\begin{tabular}{l@{\hspace{2.8mm}}l@{\hspace{2.8mm}}l@{\hspace{2.8mm}}c@{\hspace{2.8mm}}c@{\hspace{2.8mm}}c@{\hspace{2.8mm}}}
\noalign{\hrule} \noalign{\vskip3pt}
\multicolumn{6}{c}{LMC~N66}\\
$\lambda_{\rm obs}$&$F(\lambda)$&$I(\lambda)$&ID&$\lambda_{\rm 0}$&Mult  \\
\noalign{\vskip3pt} \noalign{\hrule} \noalign{\vskip3pt}

4676.74&4.67($-$1)&&O II     &4673.73&V1     \\
4679.25&4.67($-$1)&&O II     &4676.24&V1     \\
4681.15&2.74($-$1)&&N II     &4678.14&V61b   \\
4690.02&6.73($+$1)&&He II    &4685.68& 3.4   \\
4701.92&2.93($+$0)&&?         &       &       \\
4703.74&6.44($-$1)&&[Fe VII] &4698.25&       \\
\end{tabular}
\end{minipage}
\end{table}

\section{Interstellar extinction}

Before proceeding with any kind of formal analysis, nebular spectra have to be
corrected for the effects of interstellar extinction caused by dust. In nebular
studies the amount of interstellar reddening is most usually described by
\emph{c}(H$\beta$), which is the logarithmic difference between observed and
intrinsic (or else dereddened) H$\beta$ fluxes. The reddening-corrected nebular
fluxes [in units of $I$(H$\beta$) = 100] are then given by
\begin{displaymath}
I(\lambda) = 10^{c(\mathrm{H}\beta)f(\lambda)}F(\lambda),
\end{displaymath}
where \emph{f}($\lambda$) is the adopted extinction curve normalized such that
\emph{f}(H$\beta$) = 0.

A common way to derive \emph{c}(H$\beta$) in optical studies of nebulae is from
a comparison between the observed and predicted H\,{\sc i} Balmer decrement,
using the H$\alpha$/H$\beta$, H$\gamma$/H$\beta$ and H$\delta$/H$\beta$ line
ratios. This is the method we have adopted in order to deredden our optical
spectra; three preliminary \emph{c}(H$\beta$) values were obtained from each of
the above Balmer line ratios and then averaged with weights of 3 : 1 : 1,
respectively, according to the wavelength separation of the lines. The
comparison was against the theoretical ratios from Storey \& Hummer (1995) for
representative values of $T_{\rm e}$ = 10000\,K and $N_{\rm e}$ =
5000\,cm$^{-3}$. Subsequent iteration, once the true nebular temperatures and
densities were derived, resulted in negligible changes for \emph{c}(H$\beta$),
denoted as \emph{c}(H$\beta$)$^\mathrm{{Ba}}$ hereafter. The reddening
correction applied using \emph{c}(H$\beta$)$^\mathrm{{Ba}}$ effectively
compensates for errors introduced by the relative flux calibration of the
spectra. The main disadvantage over other methods is the relatively small
wavelength separation of the Balmer lines employed. The values of
\emph{c}(H$\beta$)$^\mathrm{{Ba}}$ that were used to deredden each optical
spectrum are given in Tables~3 and 4 under the heading of each PN's line-list.
The two values listed for IC\,4406 correspond to the scanned and fixed-slit
spectra, respectively.

A more accurate estimate of \emph{c}(H$\beta$) is possible through a comparison
of the observed (optically thin) radio free-free continuum radiation of
the PN, e.g. at 5\,GHz, with the observed total H$\beta$ flux. Since the
wavelength baseline involved is large and the radio flux suffers virtually no
extinction, this method gives highly reliable reddening estimates and is to be
preferred when global nebular properties are studied. In this way we derived
reddening values for our Galactic PNe using the formulae of Milne \& Aller
(1975), employing total H$\beta$ and 5\,GHz continuum fluxes taken from CKS92,
along with He$^{+}$/H$^+$ and He$^{2+}$/H$^+$ fractions deduced from our
optical observations (see Section~5.1). Similarly, one can also derive
\emph{c}(H$\beta$) from a comparison between the He~{\sc ii} $\lambda$1640 and
$\lambda$4686 or [O~{\sc ii}] $\lambda$2470 and $\lambda\lambda$7320, 7330
integrated line fluxes.

In this work we used the value of \emph{c}(H$\beta$) derived from the
radio-H$\beta$ method, denoted \emph{c}(H$\beta$)$^\mathrm{{radio}}$ (as listed
in column~3 of Table~1), to deredden the UV line fluxes in cases where the
apparent PN diameter was larger than the \emph{IUE} large aperture. All UV
fluxes were then scaled to the optical spectra as described previously in
Section~4. In our sample, apart from NGC\,5315 and the Cloud nebulae, the PNe
angular diameters are larger than the \emph{IUE} large aperture, so the derived
\emph{c}(H$\beta$)$^{\small1640}$ values are consistently greater than the one
given by \emph{c}(H$\beta$)$^\mathrm{{rad}}$, since the derivation involves a
comparison between a fraction of the He~{\sc ii} $\lambda$1640 line flux with
the total nebular He~{\sc ii} $\lambda$4686 flux; therefore
\emph{c}(H$\beta$)$^{\small1640}$ values were not used in our analysis.

The Magellanic Cloud PN spectra were dereddened in a two step process. First
the contribution from Galactic foreground reddening was removed; this was
estimated from the reddening maps of Burstein \& Heiles (1982), using the
extinction law of Howarth (1983) in all cases. The remaining extinction due to
the interstellar medium of the Magellanic Clouds was taken from Barlow (1987)
in the case of the UV data, and from the observed Balmer decrement in the case
of the optical data. The SMC extinction law of Prevot et al. (1984) was used in
the case of SMC~N87, while the mean LMC extinction law of Howarth (1983) was
used for LMC~N141. For the directions to SMC~N87 and LMC~N141,
\emph{c}(H$\beta$) = 0.03 and 0.11 through the Milky Way were found
respectively. From the foreground-corrected Balmer decrements, values of
\emph{c}(H$\beta$) = 0.02 and 0.03, respectively, were found for the remaining
Magellanic Cloud extinction. For LMC~N66, the observed optical Balmer line
ratios were already consistent with Case~B values, so no extinction correction
was applied to the optical line fluxes. We corrected the observed \emph{IUE}
line fluxes of Pena \& Ruiz (1988) by comparing the observed He~{\sc ii}
$\lambda$1640/$\lambda$4686 flux ratio of 5.1 with the theoretical Case B value
of 7.4 for {\elt} = 2$\times10^4$\,K. This yielded $c$(H$\beta$) = 0.14 (0.09
Galactic; 0.05 LMC), which was applied to deredden the UV line fluxes.

Another kind of correction had to be applied for high-excitation objects with a
high He~{\sc ii} $\lambda$4686 flux [e.g. $\gtrsim$\,65 when $F$(H$\beta$) =
100]; in such nebulae the He~{\sc ii} Pickering series (transitions
\emph{n}\,$\rightarrow$\,4) begins to contribute significantly to the H\,{\sc
i} Balmer series. Relevant cases include NGC\,2022, NGC\,2440, NGC\,6302 and
LMC~N66, where the contribution to H$\beta$ from He~{\sc ii} Pi\,8 was
estimated using the observed He~{\sc ii} $\lambda$4686 flux and emissivities
from Storey \& Hummer (1995). Fluxes of the contaminating He~{\sc ii}
$\lambda$4859 (\emph{n} = 8\,$\rightarrow$\,4) amounted to $\sim$\,3--5\,per
cent and were subtracted from the H$\beta$ fluxes before any further
analysis was performed.

Throughout the above analysis we used total nebular \emph{F}(H$\beta$) and
5\,GHz continuum fluxes for the Galactic PNe, as listed by CKS92; from them
total nebular He~{\sc ii} $\lambda$4686 fluxes were derived, using the observed
\emph{I}($\lambda$4686)/\emph{I}(H$\beta$) ratios from our scanned nebular
spectra. In the case of nebulae that were observed only with a fixed-slit we
proceeded as follows: for NGC\,6302 we used the total nebular He~{\sc ii} flux
from CKS92 directly, without resorting to our fixed-slit
\emph{I}($\lambda$4686)/\emph{I}(H$\beta$) ratio, while for NGC\,6818 the above
optical ratio was assumed to be representative of the whole PN.

\setcounter{table}{4}
\begin{table*}
\begin{minipage}{180mm}
\centering \caption{\emph{IUE} fluxes of Galactic PNe.$^a$}
\begin{tabular}{lcccccccccc}
\noalign{\vskip3pt} \noalign{\hrule} \noalign{\vskip3pt}

Line                           &\multicolumn{2}{c}{NGC\,2022}     &\multicolumn{2}{c}{NGC\,2440}         &\multicolumn{2}{c}{NGC\,3132}     &\multicolumn{2}{c}{NGC\,3242}               &\multicolumn{2}{c}{NGC\,3918}        \\
                               & $F$($\lambda$) & $I$($\lambda$)  & $F$($\lambda$) & $I$($\lambda$)      & $F$($\lambda$) & $I$($\lambda$)  & $F$($\lambda$) & $I$($\lambda$)            & $F$($\lambda$)$^b$ & $I$($\lambda$)  \\
                               & (10$^{-14}$)   & c = 0.55       & (10$^{-13}$)   & c = 0.52             & (10$^{-13}$)   & c = 0.33        & (10$^{-13}$)   & c = 0.20               & (10$^{-12}$)   & c = 0.40        \\
\noalign{\vskip3pt} \noalign{\hrule} \noalign{\vskip3pt}                                                                                                                                                                    \\
N~{\sc v}      $\lambda$1240   & *              &  *              & 28.3           &105                  &*               &*                 &*               &*                          &6.70       &41.1                  \\
O~{\sc iv}]    $\lambda$1401   & 47.8           &  124            & 21.3           &52.9                 &*               &*                 &*               &*                          &10.8       &48.9                  \\
N~{\sc iv}]    $\lambda$1486   & 38.3           &  90.1           & 82.2           &186                  &*               &*                 &*               &*                          &11.0       &46.4                   \\
C~{\sc iv}     $\lambda$1549   & 367            &  814            & 309            &659                  &49.3            &7.40              &*               &*                          &113        &458                     \\
{\fnev}        $\lambda$1574   & 9.78           &  21.3           & 5.83           &12.2                 &*               &*                 &*               &*                          & *         & *                     \\
{\fneiv}      $\lambda$1601    & 6.94           &  14.8           & *              & *                    &*               &*                 &*               &*                          &1.50       &5.90                \\
He~{\sc ii}    $\lambda$1640   & 378            &  790            & 255            &514                  &17.8            &25.9              &890             &172                        &73.3       &284                   \\
O~{\sc iii}]   $\lambda$1663   & 8.58           &  17.7           & 19.6           &39.2                 &2.08            &3.00              &43.9            &8.50                       &8.20       &31.5                   \\
N~{\sc iii}]   $\lambda$1750   & *              &  *              & 70.6           &140                  &3.18            &4.60              &28.7            &5.50                       &7.00       &26.7                   \\
C~{\sc iii}]   $\lambda$1908   & 224            &  526            & 410            &924                  &26.4            &41.1              &741             &150                        &117        &493                  \\
\noalign{\vskip3pt} \noalign{\hrule} \noalign{\vskip3pt}
Line                           &\multicolumn{2}{c}{NGC\,5315} &\multicolumn{2}{c}{NGC\,5882}         &\multicolumn{2}{c}{NGC\,6302}      &\multicolumn{2}{c}{NGC\,6818}    &\multicolumn{2}{c}{IC\,4406}           \\
                               & $F$($\lambda$)$^c$ & $I$($\lambda$)  & $F$($\lambda$) & $I$($\lambda$)      & $F$($\lambda$) & $I$($\lambda$)   & $F$($\lambda$)$^e$ & $I$($\lambda$)     & $F$($\lambda$) & $I$($\lambda$)     \\
                               & (10$^{-13}$)   & c = 0.57        & (10$^{-13}$)   & c = 0.43            & (10$^{-13}$)   &  c = 1.47        & (10$^{-12}$)   & c = 0.36           & (10$^{-13}$)   &  c = 0.30             \\
\noalign{\vskip3pt} \noalign{\hrule} \noalign{\vskip3pt}
N~{\sc v}      $\lambda$1240   & *              & *               & *              & *                   & 3.01           & 1382             & 7.90           & 15.9               & *              & *                      \\
O~{\sc iv}]    $\lambda$1401   & 3.54           & .679            & *              & *                   & 1.46           & 112              & 21.5           & 32.6               & *              & *                      \\
N~{\sc iv}]    $\lambda$1486   & 1.95           & 2.58            & *              & *                   & 6.96           & 534              & 22.2           & 31.6               & *              & *                      \\
C~{\sc iv}     $\lambda$1549   & *              & *               & *              & *                   & 8.39           & 552              & 70.8           & 96.7               & 4.84           & 19.9                   \\
{\fnev}        $\lambda$1574   & 6.21           & 7.41            & *              & *                   & *              & *                & 7.30:          & 9.84:              & *              & *                      \\
{\fneiv}      $\lambda$1601    & 2.06           & 2.41            & *              & *                   & 0.30$^d$       & 26.6             & 5.20           & 6.94               & *              & *                      \\
He~{\sc ii}    $\lambda$1640   & *              & *               & 15.3           & 17.8                & 7.24           & 409              & 242            & 318                & 21.0           & 83.8                   \\
O~{\sc iii}]   $\lambda$1663   & 3.9            & 4.42            & 4.25           & 4.90                & 2.34           & 129              & 22.8           & 29.8               & 1.97           & 7.80                   \\
N~{\sc iii}]   $\lambda$1750   & 4.7            & 5.26            & 3.90           & 4.50                & 7.61           & 408              & 16.4           & 21.2               & 1.99           & 7.80                   \\
C~{\sc iii}]   $\lambda$1908   & 24.7           & 31.9            & 24.7           & 31.4                & 6.09           & 463              & 319            & 453                & 40.6           & 140                    \\
C~{\sc ii}]    $\lambda$2326   & 6.20           & 9.46            & *              & *                   & *              & *                & 16.5           & 26.0               & 14.9           & 68.9                   \\
{\fneiv}       $\lambda$2423   & 1.72           & 1.92            & *              & *                   & 10.1$^d$       & 795              & 39.2           & 50.7               & 2.38           & 9.30                   \\
{\foii}        $\lambda$2470   & 9.05           & 8.93            & *              & *                   & *              & *                & 2.50           & 3.00               & 2.83           & 10.4                   \\

\noalign{\vskip3pt} \noalign{\hrule} \noalign{\vskip3pt}
\end{tabular}
\begin{description}
\item[$^a$] Notes: $F$($\lambda$) in units of erg\,cm$^{-2}$\,s$^{-1}$, $I$($\lambda$) in units such that $I$(\Hb) = 100.
The quoted reddening constants are from the ratio of the 5\,GHz
free--free continuum to {\Hb}. For those PNe observed at ESO with a
scanning slit the percentages of total nebular $F$($\lambda$1640) in
\emph{IUE}'s large aperture are: 27\% (NGC\,2022), 60\% (NGC\,2440), 46\%
(NGC\,3132), 53\% (NGC\,3242), 80\% (NGC\,3918), 65\% (NGC\,5882) and 48\%
(IC\,4406);
\item[$^b$] From Clegg et al. (1987) combined with fluxes
from spectrum SWP\,47514;
\item[$^c$] From Feibelman (1998);
\item[$^d$] From Feibelman (2001);
\item[$^e$] From Hyung et al. (1999).
\end{description}
\end{minipage}
\end{table*}

\section{Nebular diagnostics}

Nebular electron temperatures and densities were derived from several CEL
diagnostic ratios by solving the equations of statistical equilibrium for
multi-level ($\geq$\,5) atomic models using the program {\sc equib} (originally
written by I.D. Howarth and S. Adams). The following diagnostic ratios were
used:
\begin{itemize}
\item[]$T_{\rm e}$(O~{\sc iii}): \emph{I}($\lambda$4959 + $\lambda$5007)/\emph{I}($\lambda$4363)
\item[]$T_{\rm e}$(O~{\sc iii})$_{\rm IR}$: \emph{I}($\lambda$4959 + $\lambda$5007)/\emph{I}(52$\mu$m + 88$\mu$m)
\item[]$T_{\rm e}$(N~{\sc ii}): \emph{I}($\lambda$6548 + $\lambda$6584)/\emph{I}($\lambda$5754)
\item[]$T_{\rm e}$(O~{\sc ii}): \emph{I}($\lambda$3727)/\emph{I}($\lambda$7320 + $\lambda$7330)
\item[]$T_{\rm e}$(S~{\sc ii}): \emph{I}($\lambda$4068)/\emph{I}($\lambda$6731 + $\lambda$6716)
\item[]$T_{\rm e}$(BJ): $I_{\rm c}$($\lambda$3646$^{-}$ $-$ $\lambda$3646$^{+}$)/\emph{I}(H\,11)
\item[]$N_{\rm e}$(Ar~{\sc iv}): \emph{I}($\lambda$4740)/\emph{I}($\lambda$4711)
\item[]$N_{\rm e}$(Cl~{\sc iii}): \emph{I}($\lambda$5537/\emph{I}($\lambda$5517)
\item[]$N_{\rm e}$(S~{\sc ii}): \emph{I}($\lambda$6731)/\emph{I}($\lambda$6716)
\item[]{\eld}({\oii}): \emph{I}($\lambda$3729)/\emph{I}($\lambda$3726)
\end{itemize}

The procedure to derive temperatures and densities from CEL line ratios is as
follows: we assumed a representative initial electron temperature of 10000\,K
to derive {\eld}(Cl~{\sc iii}) and {\eld}(Ar~{\sc iv}). We then used the mean
density from these diagnostics to derive {\elt}(O~{\sc iii}) and iterated once
to get the final values. In a similar manner, {\elt}(N~{\sc ii}) was derived in
conjunction with {\eld}(O~{\sc ii}) and {\eld}(S~{\sc ii}). The atomic data
used for the purpose of this analysis, but also throughout this work, are those
used by Liu et al. (2000) in their case study of NGC\,6153.

\subsection{Electron Densities}

The derived electron densities for our sample of PNe are listed in Table~6. In
all spectra the {\fariv} $\lambda$4711 line is blended with the {\hei}
$\lambda$4713 line; individual fluxes for the two lines were thus derived by
means of Gaussian profile fitting, as described previously in Section~6. Both
of the {\fariv} $\lambda\lambda$4711, 4740 fluxes were measured on
high-resolution spectra (1.5\,{\AA} for the Galactic PNe and 2\,{\AA} for the
Magellanic Cloud PNe) in all cases; they were then used to derive the
\eld(\ariv) values listed in Table~6.

We were able to derive {\eld}'s using the {\foii} $\lambda$3729/$\lambda$3726
ratio only for NGC\,5882, 6302 and 6818, for which the {\foii}
$\lambda\lambda$3726, 3729 doublet was resolved in the high-resolution
3040--4040\,{\AA} spectra. For the remaining PNe the {\foii} doublet is covered
in lower resolution only and is blended with the H\,14 $\lambda$3721.9, [S~{\sc
iii}] $\lambda$3721.6 and H\,13 $\lambda$3734.4 lines; the overall [O~{\sc ii}]
flux was derived as follows: the [S~{\sc iii}] $\lambda$3721 flux was estimated
from a comparison with the dereddened flux of the {\fsiii} $\lambda$6312.1
line, which originates from the same upper level. The latter line is often
blended with He~{\sc ii} $\lambda$6310.8 in high excitation objects and that
flux was retrieved via its theoretical ratio relative to He~{\sc ii}
$\lambda$4686. Finally, the \emph{I}(H\,13)/\emph{I}(H$\beta$) and
\emph{I}(H\,14)/\emph{I}(H$\beta$) intensity ratios were estimated using
H\,{\sc i} line emissivities from Storey \& Hummer (1995). The deconvolved
$\lambda$3727 flux was then used to derive {\eld}'s for the remaining PNe of
Table~6 from the {\foii} $I$($\lambda$3727)/$I$($\lambda$7320 + $\lambda$7330)
ratio.

Electron densities were also derived from the {\foiii} 52\,$\mu$m/88\,$\mu$m
\emph{ISO} LWS line flux ratios measured for eight of our sample PNe by Liu et
al. (2001a)\footnote{Liu et al. adopted a constant electron temperature of
10$^4$\,K to derive electron densities from their 52\,$\mu$m/88\,$\mu$m flux
ratios and compared their values to those derived from the [Cl~{\sc iii}] and
[Ar~{\sc iv}] optical doublet ratios.}.

\setcounter{table}{5}
\begin{table*}
\begin{minipage}{100mm}
\centering \caption{Electron densities.}
\begin{tabular}{l@{\hspace{3mm}}c@{\hspace{3mm}}c@{\hspace{3mm}}c@{\hspace{3mm}}c@{\hspace{3mm}}c@{\hspace{3mm}}}
\noalign{\vskip3pt} \noalign{\hrule} \noalign{\vskip3pt}
PN &[O~{\sc ii}]$^a$  &[S~{\sc ii}] &[Cl~{\sc iii}] &[Ar~{\sc iv}]  &{\foiii}$^b$\\
   &$\lambda$3727/($\lambda$7320 + $\lambda$7330)  &$\lambda$6731/$\lambda$6716
&$\lambda$5537/$\lambda$5517 &$\lambda$4740/$\lambda$4711  &52\,$\mu$m/88\,$\mu$m  \\
&\multicolumn{5}{c}{$N_{\rm e}$~(cm$^{-3}$)} \\
\noalign{\vskip3pt} \noalign{\hrule} \noalign{\vskip3pt}
NGC\,2022   &1970       &1050       &850     &2150     &*\\
\noalign{\vskip2pt}
NGC\,2440   &*      &*          &*        &4000                         &*\\
\noalign{\vskip2pt}
NGC\,3132   &*        &550       &720       &530         &355\\
\noalign{\vskip2pt}
NGC\,3242   & *     &1970    &1200     &3040        &775 \\
\noalign{\vskip2pt}
NGC\,3918   & *     &4600    &5500     &6900           &1380 \\
\noalign{\vskip2pt}
NGC\,5315   &13040    &8200    &22825      &12300     &2290 \\
\noalign{\vskip2pt}
NGC\,5882   &4750     &4000    &2700     &5000     &1175\\
\noalign{\vskip2pt}
NGC\,6302   &5750    &12900    &22450     &14900       &1380 \\
\noalign{\vskip2pt}
NGC\,6818   &1800     &1700    &2400     &2350         &*\\
\noalign{\vskip2pt}
IC\,4191$^c$   &13530     &12750       &12375     &13750  &*\\
\noalign{\vskip2pt}
IC\,4191$^d$   &11930     &7900       &10150     &12800   &1700\\
\noalign{\vskip2pt}
IC\,4406    & *     &950     &3500      &1250          &540\\
\noalign{\vskip2pt}
My\,Cn\,18   & *   &5025    &9420     &6300            &*\\
\noalign{\vskip2pt}
SMC\,N87    &2850  &3950       & *         &9500          &*\\
\noalign{\vskip2pt}
LMC\,N66    & $\leq$200   &1900        & *      &5700         &*\\
\noalign{\vskip2pt}
LMC\,N141   &2300    &7400   & *         &9500            &*\\

\noalign{\vskip3pt} \noalign{\hrule}
\end{tabular}
\begin{description}
\item[$^a$] For NGC\,5882, 6302, 6818 and LMC~N66 quoted values were derived
from the $\lambda$3729/$\lambda$3726 ratio;
\item[$^b$] From the \emph{ISO} LWS line-fluxes of Liu et al. (2001);
\item[$^c$] Values from the fixed-slit spectrum of the nebula;
\item[$^d$] Values for the whole nebula from a scanned spectrum.
\end{description}
\end{minipage}
\end{table*}

It should be borne in mind that the derived densities correspond to mean
values, since the employed line ratios have been taken from scanned nebular
spectra in most cases. In addition, all observations sampled lines of sight
throughout the nebular volumes as well. It is thus quite probable that
localized density variations on a small scale have been effectively smoothed
out, or that they are not even seen due to the relatively low spatial
resolution of our observations.

On the other hand, the {\eld}(\sii) values for the whole sample are
consistently lower than the {\eld}(\ariv) values, by about $\sim$30\,per cent.
This behaviour is consistent with the presence of strong density variations in
the nebulae, so that the diagnostic line ratios with higher critical densities
(e.g. the {\fcliii} and {\fariv} diagnostics) yield higher derived nebular
electron densities.\footnote{{\fcliii} $\lambda\lambda$5517, 5537, {\crd} =
6400, 34\,000\,{\cmt}; {\fariv} $\lambda\lambda$4711, 4740, {\crd} = 14\,000,
130\,000\,{\cmt}; {\fsii} $\lambda\lambda$6716, 6730, {\crd} = 1200,
3300\,{\cmt} respectively.} The effect is shown to be more pronounced from a
comparison between the {\foiii} 52-$\mu$m/88-$\mu$m and {\fcliii} and {\fariv}
densities. For the eight PNe for which values from all three of
these ratios have been
derived, the latter diagnostics yield {\eld}'s that are on average a factor of
6 higher than those obtained from the far-IR ratio (see Rubin 1989, Liu et al.
2001a). This effect will be discussed further in Section~5.2.

\subsection{Electron temperatures}

We derived {\elt}'s using ionic CEL ratios of nebular to transauroral {\fsii}
transitions, nebular to auroral transitions of {\fnii}, {\foii} and {\foiii},
and nebular to infrared fine-structure transitions of {\foiii} (the latter
designated as $T_{\rm e}$(O~{\sc iii})$^{\rm IR}$). We adopted the ISO Long
Wavelength Spectrometer (LWS) fluxes listed by Liu et al. (2001) for the
{\foiii} 52~$\mu$m and 88~$\mu$m lines. The LWS aperture size of
$\sim$80~arcsec included all of the emission from these nebulae. The results
are presented in Table~7. We were able to derive values of \elt(\oiii) for all
15 PNe; values of \elt(\nii) for 13 PNe, excluding SMC~N87 and LMC~N141 for
which the auroral $\lambda$5754 line did not fall within our wavelength
coverage; and values of \elt(\oiii)$^{\rm IR}$ for eight PNe. The correct
choice of electron temperature is crucial when deriving abundances from
forbidden lines, so in what follows we discuss the pattern of \elt's for the
current sample of 15 PNe, as measured from the above ionic ratios.

\setcounter{table}{6}
\begin{table*}
\centering \caption{Electron temperatures.}
\begin{tabular}{l@{\hspace{6mm}}c@{\hspace{6mm}}c@{\hspace{6mm}}c@{\hspace{6mm}}c@{\hspace{6mm}}c@{\hspace{6mm}}c@{\hspace{6mm}}c@{\hspace{6mm}}c@{\hspace{6mm}}}
\noalign{\vskip3pt} \noalign{\hrule} \noalign{\vskip3pt}
PN &[S~{\sc ii}]  &[O~{\sc ii}] &[N~{\sc ii}] &[O~{\sc iii}]  &BJ
&[O~{\sc iii}]$^{\rm IR}$     &[O~{\sc iii}]$^{\rm IR}_{\rm Cl~III}$    &{\oii} ORLs$^a$  \\
&\multicolumn{7}{c}{$T_{\rm e}$~(K)} & \\
 \noalign{\vskip3pt} \noalign{\hrule} \noalign{\vskip3pt}

NGC 2022   & *        & *        &14700     &15000   &13200   &*   &*  &400\\
\noalign{\vskip2pt}
NGC 2440   &*        & *        &*         &16150     &14000      &*  &*  &*\\
\noalign{\vskip2pt}
NGC 3132   &8120      & *           &9350       &9530 &10780      &9900  &9000    &*\\
\noalign{\vskip2pt}
NGC 3242   &4800     &19550    &13400     &11700     &10200   &13800  &12400  &1000\\
\noalign{\vskip2pt}
NGC 3918   &9350     &10400    &10800     &12600     &12300   &19900  &12400   &*\\
\noalign{\vskip2pt}
NGC 5315   &11400    & *     &10800      &9000       &8600    &18500  &7800    &8100\\
\noalign{\vskip2pt}
NGC 5882   &6900     &15300    &10800     &9400      &7800    &11400  &9400   &800\\
\noalign{\vskip2pt}
NGC 6302   &10000    &20200    &14225     &18400     &16400       &$>$50000 &10900   &*\\
\noalign{\vskip2pt}
NGC 6818   &5750     &11200    &11100     &13300     &12140      &*  &*   &$\lesssim$2900\\
\noalign{\vskip2pt}
IC 4191$^b$   &7750     & *        &11575     &10700    &10500   & *  &* &*\\
\noalign{\vskip2pt}
IC 4191$^c$   &8550     & *        &12225     &10000     &9200   &16700 &8900   &4300  \\
\noalign{\vskip2pt}
IC 4406    &8650     &9000     &9900      &10000     &9350       &9900 &7100    &*\\
\noalign{\vskip2pt}
My\,Cn 18   &12825    &12300    &10225     &7325      &*          &* &*       &*\\
\noalign{\vskip2pt}
SMC N87    &$>$20000   & *        & *        &12250     &*        &* &*&*\\
\noalign{\vskip2pt}
LMC N66    & *        &*       &12700     &18150     &*         &*&*&*\\
\noalign{\vskip2pt}
LMC N141   &10100     & *        & *        &11850     &*         &* &*&*\\

\noalign{\vskip3pt} \noalign{\hrule}
\end{tabular}
\begin{description}
\item[$^a$] {\oii} ORL temperatures from Paper~II
\item[$^b$] Values from a fixed-slit spectrum of the nebula;
\item[$^c$] Values for the whole nebula from a scanned spectrum.
\end{description}
\end{table*}

\subsubsection{Recombination excitation of the {\nii} and {\oii} auroral lines}

The mean \elt(\oiii) for the whole sample is 12300\,K, while the mean
\elt(\nii) is 11700\,K. For 12 PNe where both \elt's were measured, the values
differ by about 3\,per cent, \elt(\oiii) being only slightly higher than
\elt(\nii). It could be argued that generally in PNe, N$^+$ is present in a
lower ionization zone than O$^{2+}$, and that the temperature should be lower
too. However, for 8 PNe where both \elt(\oiii) and \elt(\oii) can be measured,
it is found that on average, \elt(\oiii) = 11500\,K, while \elt(\oii) =
13400\,K, i.e. \emph{higher} by 1900\,K. The disagreement between {\foii} and
{\foiii} electron temperatures, as derived from the nebular to auroral line
ratios $I(\lambda3727)/I(\lambda7320+\lambda7330)$ and
$I(\lambda4959+\lambda5007)/I(\lambda4363)$, respectively, is at its most
extreme in the cases of NGC\,3242, NGC\,5882 and My\,Cn\,18. In these objects
the derived \elt(\oii)'s are higher than the \elt(\oiii)'s by 7850\,K, 5900\,K
and 4980\,K, respectively.

A variety of hypotheses can be invoked in order to explain the difference of
electron temperatures in these three nebulae. Radiation hardening in the
nebular volumes where singly ionized species are expected to exist, could
result in higher temperatures than those of the O$^{2+}$ zone, but the effect
should be similar for both \elt(\nii) and \elt(\oii), since the ionization
potentials of N$^+$ and O$^+$ are very comparable.

An alternative proposition is as follows: in NGC\,3242, NGC\,5882 and
My\,Cn\,18 most of the N and O atoms are in their doubly ionized stages
(O$^{2+}$/O = 0.85, 0.96 and 0.63, while N$^{2+}$/N = 0.65, 0.72 and 0.63,
respectively; cf. Section~5). Thus, as discussed by Rubin (1986) and Liu et al.
(2000), recombination of N$^{2+}$ and O$^{2+}$ can be important in contributing
to the excitation of the {\fnii} nebular and {\foii} nebular and auroral lines.

We will attempt to quantify this effect for these three extreme cases, using
the expressions of Liu et al. (2000). They employed radiative recombination
coefficients from P\'{e}quignot, Petitjean \&Boisson (1991) and dielectronic
recombination coefficients from Nussbaumer \& Storey (1984), to show that the
intensity of the {\fnii} $\lambda$5754 auroral line due to recombination
excitation can be fitted by

\begin{equation}\label{RECEQ1}
\frac{I_{\rm R}(\lambda5754)}{I({\rm H}\beta)} =
3.19\,t^{0.30}\,\times\,\frac{~\rm N^{2+}}{\rm H^{+}},
\end{equation}
where \emph{t}\,$\equiv$\,\elt/10$^4$ and 0.5\,$\leq$\,\emph{t}\,$\leq$\,2.0.
Also, Liu et al. (2000) calculated new recombination coefficients for the
metastable levels of O~{\sc ii} and
showed that the intensity of the {\foii} $\lambda\lambda$7320, 7330
auroral lines due to recombination excitation can be fitted in the range
0.5\,$\leq$\,\emph{t}\,$\leq$\,1.0 by

\begin{equation}\label{RECEQ2}
\frac{I_{\rm R}(\lambda7320\,+\,\lambda7330)}{I({\rm H}\beta)} =
9.36\,t^{0.44}\,\times\,\frac{~\rm O^{2+}}{\rm H^{+}}.
\end{equation}

For NGC\,3242, which has {\elt}(\oiii) = 11700\,K, the observed $\lambda$1750
multiplet flux yields N$^{2+}$/H$^+$ = 1.90\,$\times$\,10$^{-5}$, for \elt
= 11700\,K and
\eld = 1200\,{\cmt} (Tables~6, 7). Inserting this value into Eq.\,~1, we have
$I_{\rm R}(\lambda5754)/I({\rm H}\beta)$ = 0.0064, or 10\,per cent of the
observed intensity of the $\lambda$5754 line relative to \Hb,
$I(\lambda5754)/I$(\Hb) = 0.0635. After subtracting $I_{\rm R}(\lambda5754)$
from the observed intensity, the {\fnii} nebular to auroral line ratio yields
{\elt} = 10\,850\,K, i.e. 2550\,K lower than the value derived before the
correction. If one uses instead N$^{2+}$/H$^+$ = 1.28\,$\times$\,10$^{-4}$, as
derived from {\nii} recombination lines (cf. Paper~II for {\npp}/{\hp} and
{\opp}/{\hp} abundances from {\nii} and {\oii} ORLs), one gets $I_{\rm
R}(\lambda5754)/I({\rm H}\beta)$ = 0.0408, which is about 64\,per cent of the
observed value. After correcting for this $I_{\rm R}(\lambda5754)$, the
temperature deduced from the revised {\fnii} line ratio is only 7950\,K.

Regarding the {\foii} auroral lines, for O$^{2+}$/H$^+$ =
2.80\,$\times$\,10$^{-4}$ as derived from the $\lambda\lambda$4959, 5007 lines
(Table~9), Eq.\,~2 predicts a recombination intensity relative to {\Hb} of
0.262 on a scale where {\Hb} = 100 for the {\foii} $\lambda\lambda$7320, 7330
lines, or 29\,per cent of their observed intensity. According to Liu et al.,
the recombination contribution to the $\lambda\lambda$3726, 3729 nebular lines
is 7.7 times that of $I_{\rm R}(\lambda7320+\lambda7330)$, at {\elt} =
10\,000\,K. Thus we have, $I_{\rm R}(\lambda3726+\lambda3729)$ = 2.02 on a
scale where {\Hb} = 100, or 21\,per cent of the observed overall intensity.
Again, if we use instead the ORL abundance ratio of O$^{2+}$/H$^+$ =
6.28\,$\times$\,10$^{-4}$ derived for this PN, we get a recombination
contribution of 0.588 to the {\foii} auroral lines (65\,per cent of the
observed value), while a contribution of 4.53 relative to {\Hb} (46\,per cent
of the observed intensity), is found for the nebular lines.

Therefore, using O$^{2+}$/H$^+$ as given by the CEL lines, the revised {\foii}
nebular to auroral line ratio yields {\elt} = 15800\,K, i.e. 3750\,K lower
than the value derived before the correction. If we instead use the
{\opp}/{\hp} abundance ratio given by the ORLs, the corrected {\elt}(\oii) is
7500\,K lower, at 12050\,K, and only 300\,K higher than the {\elt}(\oiii).

Applying the same procedure to NGC\,5882, which has {\elt}(\oiii) = 9400\,K, we
obtain from N$^{2+}$/H$^+$ = 1.10\,$\times$\,10$^{-4}$, as given by the CELs, a
corrected nebular to auroral line ratio which yields a {\elt}(\nii) of 9900\,K;
if we assume N$^{2+}$/H$^+$ = 1.63\,$\times$\,10$^{-4}$, as given by the ORLs,
the corrected temperature is {\elt}(\nii) = 9550\,K. Similarly, the corrected
{\foii} nebular to auroral line ratio results in a {\elt}(\oii) of 14500\,K,
using O$^{2+}$/H$^+$ = 4.48\,$\times$\,10$^{-4}$ as given by CELs; in the case
of O$^{2+}$/H$^+$ = 9.70\,$\times$\,10$^{-4}$, as given by ORLs, the corrected
temperature is {\elt}(\oii) = 13750\,K.

For My\,Cn\,18, which has {\elt}(\oiii) = 7325\,K, the {\fnii} line ratio does
not change appreciably for N$^{2+}$/H$^+$ = 2.20\,$\times$\,10$^{-4}$ as given
by CELs, so there is a negligible change in {\elt}(\nii); however if
N$^{2+}$/H$^+$ = 20.4\,$\times$\,10$^{-4}$, as given by the ORLs, the corrected
temperature becomes {\elt}(\nii) = 9450\,K. Finally, the corrected {\foii}
ratio results in a {\elt}(\oii) of 12150\,K, using O$^{2+}$/H$^+$ =
3.54\,$\times$\,10$^{-4}$ as given by CELs; and for O$^{2+}$/H$^+$ =
6.43\,$\times$\,10$^{-4}$, as given by ORLs for this object, the corrected
temperature is {\elt}(\oii) = 12100\,K.

In the above discussion we have neglected the contribution of recombination
excitation to the observed {\fnii} $\lambda\lambda$6548, 6584 nebular
intensities; this is estimated to be small and amounts to only 7, 8 and 8\,per
cent, respectively, for NGC\,3242, NGC\,5882 and My\,Cn\,18, even when adopting
ORL N$^{2+}$/H$^+$ abundances. We base this on the fact that for pure
recombination excitation, the {\fnii} nebular to auroral line ratio has a value
of 5.9 at {\elt} = 10\,000\,K (Liu et al. 2000). On the other hand, the
observed ratios for these three PNe are respectively, 10, 12 and 13 times
larger than that. We summarize our findings for these three objects in
Table~8.

To conclude, for low-density uniform nebular media, as assumed above,
appreciable fractions of the {\foii} nebular and auroral and {\fnii} auroral
lines can be accounted for by recombination processes. However, even after
recombination excitation has been taken into account, the agreement between the
various CEL temperature diagnostics remains quite poor, especially for
NGC\,5882 and My\,Cn\,18 (see Table~8). In addition, the current treatment has
not taken into account collisional de-excitation of the metastable levels that
are populated by recombination. It is obvious that an exact analysis of this
problem requires detailed knowledge of the actual electron temperature and
N$^{2+}$/H$^+$ and O$^{2+}$/H$^+$ abundances. To complicate matters more, if
the nebulae are not homogeneous, but also contain a higher-density component,
then the observed {\foii} and {\fnii} emission pattern can be more difficult to
evaluate. In nebulae that contain condensations whose electron density is
\emph{higher} than the critical densities of the low-lying atomic levels---from
which the $\lambda\lambda$6548, 6584 and 3727 nebular lines originate---but
\emph{lower} than that of the auroral lines, the former lines will be
preferentially emitted from the lower density medium. This could lead to
apparently high temperatures as deduced from the {\fnii} and {\foii} nebular to
auroral ratios, since their component transitions would be disproportionately
affected in regions of high density; there, the nebular lines will be
suppressed by collisional de-excitation, but the auroral lines will not be
(Viegas \& Clegg 1994).

\setcounter{table}{7}
\begin{table}
\begin{center}
\footnotesize \caption{{\fnii} and {\foii} electron temperatures after
correcting for the effects of recombination excitation.} \label{TRECEXPN}
\begin{tabular}{l@{\hspace{2.2mm}}r@{\hspace{2.2mm}}r@{\hspace{2.2mm}}r@{\hspace{2.2mm}}r@{\hspace{2.2mm}}}
\noalign{\vskip3pt} \noalign{\hrule} \noalign{\vskip3pt}
PN    &{\elt}(\oiii) &{\elt}(BJ) &{\elt}$^{\rm cor}$(\nii)  &{\elt}$^{\rm cor}$(\oii) \\
&\multicolumn{4}{c}{(K)}\\
\noalign{\vskip3pt} \noalign{\hrule} \noalign{\vskip3pt}
NGC\,3242      &11700  &10200    &10850$\mid$7950$^a$      &15800$\mid$12050$^a$   \\
NGC\,5882      &9400     &7800       &9900$\mid$9550         &14500$\mid$13750   \\
My\,Cn\,18     &7325     &*          &10225$\mid$9450     &12150$\mid$12100   \\
\noalign{\vskip3pt} \noalign{\hrule} \noalign{\vskip3pt}
\end{tabular}
\begin{description}
\item[$^a$] Values before the vertical dashes are where recombination
excitation contributions were calculated adopting CEL abundances; those
after are where the ORL abundances from Paper~II were adopted.
\end{description}
\end{center}
\end{table}

As our schematic treatment has pointed out, the {\fnii} and {\foii} nebular to
auroral temperature diagnostic ratios provide values which are probably poor
indicators of {\elt} in the low-ionization regions of these three objects.
Therefore, in the abundance analysis that follows in Section~7, we have adopted
{\elt}(\oiii) when calculating forbidden-line N$^{+}$/H$^+$ and
O$^{+}$/H$^+$ abundances for NGC\,3242, NGC\,5882 and My\,Cn\,18.

\subsubsection{Balmer discontinuity electron temperatures}

Apart from temperatures derived from the CEL ratios, Table~7 also lists the
mean nebular Balmer jump temperatures, {\elt}(BJ), derived from the ratio of
the nebular continuum discontinuity at 3646\,{\AA} to H\,11 $\lambda$3770
[$\Delta$(BJ)/H\,11]. Values of the Balmer jump intensity were determined from
the lower resolution 4.5\,{\AA} FWHM scanned spectra, apart from the cases of
NGC\,5882, 6302 and 6818 for which their higher resolution (1.5\,{\AA} FWHM)
$\lambda\lambda$3040--4040 fixed-slit spectra were used. The observed nebular
continuum on either side of the Balmer discontinuity was modelled with
low-order polynomials. Redwards of 3646\,{\AA} the crowding of high-order
nebular Balmer lines frustrates the determination of the local continuum level
on the lower resolution spectra, which was then estimated by linear
extrapolation using continuum data points with wavelengths longer than
3800\,{\AA}. In the higher resolution spectra the high-order Balmer lines are
clearly resolved and the local continuum is better estimated with no
significant extrapolation from longer wavelengths. In practice, the Balmer jump
temperatures were derived by comparing the observed and predicted values of the
Balmer discontinuity to H\,11 ratio, defined as $\Delta$(BJ)/H$\beta$ $\equiv$
[$I_{\rm c}(\lambda3643) - I_{\rm c}(\lambda3681)$]/$I$(H\,11). By thus
defining the Balmer discontinuity we include in our computation two weak
discontinuities of the He\,{\sc i} and He~{\sc ii} continua which are present
at 3678 and 3646\,{\AA}, respectively and are inseparable from the H\,{\sc i}
Balmer jump at the resolution of our observations. By definition then, the
temperature thus deduced has a weak dependence on the He$^+$/H$^+$ and
He$^{2+}$/H$^+$ abundance ratios (Table~9).  Equation (3) of Liu et al. (2001b)
takes into account the Balmer Jump dependence on these abundance ratios and we
have used it to derive the Balmer Jump temperatures listed in Table~7.

Inspection of Table~7 shows that, apart from NGC~3132, all of the Balmer
Jump temperatures in column~6 of Table~7 are lower than the corresponding
optical {\foiii} temperatures listed in column~5 (10\% lower on average).
This behaviour has been found from previous observations too (e.g. Peimbert
1971 and Liu \& Danziger 1993) and has often been attributed to nebular
temperature fluctuations, since the optical {\foiii} lines, whose
strengths have an exponential dependence on electron temperature, will be
more biased towards regions with higher than average $T_e$ than will the
hydrogen Balmer series lines and discontinuity, which have only an inverse
power-law dependence of $T_e$. However, the strengths of the far-infrared
fine-structure (FS) lines of {\foiii} have an even weaker dependence on
$T_e$, because of their low excitation energies. Thus, if
significant temperature fluctuations are present, electron temperatures
derived from the ratio of {\foiii} infrared FS lines to optical nebular
lines might be expected to show less bias to higher temperatures than those
derived from the more temperature-sensitive auroral to nebular line ratio.
In the next section, we will investigate whether this expectation is
confirmed by the observations.


\subsubsection{Electron temperatures using the far-infrared [O~{\sc iii}]
52- and 88-$\mu$m lines}

We have also derived [O~{\sc iii}] electron temperatures using the far-infrared
fine structure lines and optical nebular lines of this species, in a similar
manner to Dinerstein, Lester \& Werner (1985). These  $T_e$'s are designated as
$T_e$(O~{\sc iii})$^{\rm IR}$ in Table~7 and were obtained from the dereddened
flux ratio $I$($\lambda$4959 + $\lambda$5007)/$I$(52\,$\mu$m + 88\,$\mu$m)
using our scanned-slit optical line fluxes and the infrared line fluxes of Liu
et al. (2001), for the eight Galactic PNe in our sample that were observed by
the ISO LWS. Two alternative values of $T_e$(O~{\sc iii})$^{\rm IR}$ are listed
for each of the eight nebulae. The first value, listed in column~7 of Table~7,
was obtained using the [O~{\sc iii}] 52\,$\mu$m/88\,$\mu$m electron density
listed in column~6 of Table~6, while the second value, designated $T_e$(O~{\sc
iii})$^{\rm IR}_{\rm Cl~III}$ and listed in column~8 of Table~7, was derived
using the (higher) optical [Cl~{\sc iii}] electron densities listed in column~4
of Table~6.

Inspection of Table~7 shows that, apart from IC~4406, the $T_e$(O~{\sc
iii})$^{\rm IR}$ values derive using [O~{\sc iii}] 52\,$\mu$m/88\,$\mu$m
electron densities are always higher (often much higher) than the purely
optical $T_e$(O~{\sc iii}) values listed in column~5 of Table~7.  On average,
$T_e$(O~{\sc iii})$^{\rm IR}$ based on the 52$\mu$m/88$\mu$m electron density
is 40\% higher than the purely optical $T_e$(O~{\sc iii}). In contrast to this,
$T_e$(O~{\sc iii})$^{\rm IR}$ based on the optical [Cl~{\sc iii}] electron
density shows much closer agreement with the optical $T_e$(O~{\sc iii}). For
one nebula (NGC~3242), this $T_e$(O~{\sc iii})$^{\rm IR}$ is higher than the
optical $T_e$(O~{\sc iii}) by 6\%, while for NGC~5882 the two values agree
exactly. For the other six nebulae, the $T_e$(O~{\sc iii})$^{\rm IR}$ values
based on [Cl~{\sc iii}] electron densities are lower than the purely optical
$T_e$(O~{\sc iii}) values by only 14\% on average. We note that the trend from
our ISO-based results disagrees with that found by Dinerstein et al. (1985),
who used 52- and 88-$\mu$m fluxes and electron densities from KAO spectra of
six PNe to derive $T_e$(O~{\sc iii})$^{\rm IR}$ values that were systematically
{\em lower} than $T_e$'s derived from auroral to nebular {\foiii} line ratios.

Our results indicate that the observed $I$($\lambda$4959 +
$\lambda$5007)/$I$(52\,$\mu$m + 88\,$\mu$m) ratios are much more weighted to
the higher electron densities diagnosed by the optical [Cl~{\sc iii}]
$\lambda$5517/$\lambda$5537 ratio than to the lower electron densities
diagnosed by the 52\,$\mu$m/88\,$\mu$m ratio. In most cases, an electron
density just below that from the [Cl~{\sc iii}] ratio would yield agreement
between the optical and infrared $T_e$(O~{\sc iii}) values. We note that if
significant nebular temperature fluctuations were the cause of the Balmer jump
electron temperatures being lower than the optical [O~{\sc iii}] temperatures
(Section 4.2.2), we would expect [O~{\sc iii}] temperatures that made use of
the more temperature insensitive 52- and 88-$\mu$m lines to be significantly
lower than [O~{\sc iii}] temperatures based on the much more temperature
sensitive auroral $\lambda$4363 line. This is not the case -- the [O~{\sc
iii}]$^{\rm IR}$ temperatures based on 52\,$\mu$m/88\,$\mu$m electron densities
are significantly {\em higher}, while [O~{\sc iii}]$^{\rm IR}$ temperatures
based on [Cl~{\sc iii}] electron densities are close to, or only slightly
smaller, than the optical $T_e$(O~{\sc iii}) values. We therefore conclude that
nebular temperature fluctuations are not a significant contributor to the
trends shown by the various $T_e$ diagnostics.

\section{Ionic and total elemental abundances}

\subsection{Helium abundances}

Helium abundances derived from {\hei} and {\heii} recombination lines are given
in Tables~9 and 10. For the {\hei} lines, Case A recombination was assumed for
the triplet lines $\lambda$4471, $\lambda$5876 and Case B for the singlet
$\lambda$6678 line. The adopted effective recombination coefficients were from
Brocklehurst (1972). Contributions to the observed fluxes by collisional
excitation from the He$^{\rm 0}$ 2s\,$^3$S metastable level by electron impacts
were corrected for using the formulae derived by Kingdon \& Ferland (1995a).
The He$^+$/H$^+$ abundance ratios thus derived were then averaged, weighted
according to the intensities of the three He~{\sc i} lines. The tabulated
He$^{2+}$/H$^+$ ratios were obtained from the He~{\sc ii} $\lambda$4686 line
only, using the effective recombination coefficients of Storey \& Hummer
(1995). The tabulated total He/H number ratio for each nebula was given by He/H
= He$^+$/H$^+$ + He$^{2+}$/H$^+$.

\subsection{Abundances from collisionally excited lines}

In this section we present our results for the heavy element ionic abundances
of the 12 Galactic and 3 Magellanic Cloud PNe that were observed, derived from
collisionally excited UV, optical and IR lines. These are presented in Tables~9
and 10 for the Galactic and Cloud nebulae respectively. The emission lines
detected by the \emph{IUE} satellite gave us access to important ionic stages
such as: C$^+$--\,C~{\sc ii}] $\lambda$2326, C$^{2+}$--\,C~{\sc iii}]
$\lambda$1908, and C$^{3+}$--\,C~{\sc iv} $\lambda$1550 (resonantly excited
doublet line); N$^{2+}$--\,N~{\sc iii}] $\lambda$1750, N$^{3+}$--\,N~{\sc iv}]
$\lambda$1486, and N$^{4+}$--\,N~{\sc v} $\lambda$1240 (resonant doublet);
O$^+$--\,[O~{\sc ii}] $\lambda$2470, O$^{2+}$--\,O~{\sc iii}] $\lambda$1663,
and O$^{3+}$--\,O~{\sc iv} $\lambda$1401; Ne$^{3+}$--\,[Ne~{\sc iv}]
$\lambda\lambda$1601, 2423, and Ne$^{4+}$--\,Ne~{\sc v} $\lambda$1574.

Especially with regard to carbon, the consideration of UV lines facilitated the
calculation of forbidden line abundances for this element, which has no
emission lines in the optical domain other than recombination lines (of
C$^{2+}$, C$^{3+}$ and C$^{4+}$). The nitrogen abundance estimate also benefits
greatly from the inclusion of UV lines in the analysis, since in the optical
region---in terms of CELs---this important element is represented only by the
[N~{\sc ii}] $\lambda\lambda$5754,\,6548, 6584 lines, which are typically
indicative of just a small fraction of the total N abundance.

A complementary analysis of the more frequently employed optical CELs allowed
us to obtain total CEL elemental abundances for C, N, O, Ne, S, Cl, and Ar.

In order to derive ionic abundances we used the {\sc equib} code, which solves
the statistical balance equation for each ion and yields level populations and
line emissivities for a specified (\elt,\,\eld), appropriate to the zones in a
nebula where the ions are expected to exist. The following expression was then
used in order to convert the observed line intensities to ionic abundance
fractions;

\begin{displaymath}
\frac{~N(\rm X^{m+})}{N(\rm H^{+})} = \frac{~N_{\rm e}\,h\nu_{42}\,\alpha_{\rm
eff}(\rm H\beta)} {A_{ij}\,n_{i}\,E_{ij}}~\frac{I(\lambda)}{I(\rm H \beta)}
\end{displaymath}
where \emph{I}($\lambda$) are dereddened line fluxes, $A_{ij}$ are the
radiative transition probabilities for \emph{i}$\rightarrow$\emph{j}, $n_{i}$
is the fractional population of level \emph{i}, $E_{ij}$ is the excitation
energy of level \emph{i} above \emph{j}, $\alpha_{\rm eff}$(\Hb) is the
effective recombination coefficient of {\Hb} and $h\nu_{42}$ is the energy of
an {\Hb} photon.

Generally when deriving CEL abundances, \elt(\nii) is assumed to represent the
electron temperature appropriate for singly ionized species, while \elt(\oiii)
is used for higher excitation ions. However, as discussed in Section~4.2.1, in
some cases the {\fnii} and {\foii} nebular to auroral line ratios can be
affected in ways that result in them being unreliable diagnostics of the
temperature pertinent to the lower excitation zones of PNe. Therefore, for our
CEL abundance analysis we have adopted the following scheme: abundances of
singly ionized species were derived using {\elt}(\nii) and {\eld}(\oii), apart
from the cases of NGC\,3242, NGC\,5882 and My\,Cn\,18, where {\elt}(\oiii) was
used instead; whenever {\eld}(\oii) was not available, {\eld}(\sii) was
employed. For all doubly ionized species (C$^{2+}$, N$^{2+}$, O$^{2+}$,
Ne$^{2+}$, S$^{2+}$, Cl$^{2+}$ and Ar$^{2+}$), {\elt}(\oiii) and {\eld}(\cliii)
were used. For triply ionized species (O$^{3+}$, N$^{3+}$, Ne$^{3+}$,
Ar$^{3+}$), {\elt}(\oiii)\,+\,1000\,K with {\eld}(\ariv) were used, except for
C$^{3+}$, where {\elt}(\oiii)\,+\,650\,K was used instead. Finally,
{\elt}(\oiii)\,+\,2270\,K, was used for all four-times ionized species
(N$^{4+}$, Ne$^{4+}$, Ar$^{4+}$)---from ionization potential considerations as
discussed by Kingsburgh \& Barlow (1994; hereafter KB94).

\setcounter{table}{8}
\begin{table*}
\centering
\begin{minipage}{125mm}
\caption{Helium abundances derived from optical recombination lines and CEL
O, N, C, Ne, Ar, S and Cl abundances relative to hydrogen in Galactic PNe.$^a$}
\begin{tabular}{l@{\hspace{3mm}}l@{\hspace{3mm}}c@{\hspace{3mm}}c@{\hspace{3mm}}c@{\hspace{3mm}}c@{\hspace{3mm}}c@{\hspace{3mm}}c@{\hspace{3mm}}c@{\hspace{3mm}}}
 \noalign{\vskip3pt} \noalign{\hrule}  \noalign{\vskip3pt}
$\lambda_0$\,({\AA})     &           &NGC\,2022  &NGC\,2440  &NGC\,3132   &NGC\,3242  &NGC\,3918  &NGC\,5315   \\
\noalign{\vskip3pt} \noalign{\hrule} \noalign{\vskip3pt}
4471      &  He$^{+}$/H$^{+}$        &0.0135     &0.0575     &0.1229      &0.0811     &0.0721     &0.1243          \\
5876      &  He$^{+}$/H$^{+}$        &0.0129     &    *      &0.1151      &0.0789     &0.0655     &0.1205          \\
6678      &  He$^{+}$/H$^{+}$        &*          &    *      &0.1103      &0.0762     &0.0631     &0.1174       \\
Avg.      &  He$^{+}$/H$^{+}$        &0.0130     &0.0575     &0.1157      &0.0789     &0.0663     &0.1206       \\
4686      &  He$^{2+}$/H$^{+}$       &0.0970     &0.0627     &0.0031      &0.0208     &0.0350     &0.0000          \\
          &  He/H                    &0.110      &0.120      &0.119       &0.100      &0.101      &0.121        \\
2326 &          C$^{+}$/H$^{+}$      &  *        &   *       &*           & *         &3.73($-$5) &1.54($-$5)         \\
1908 &          C$^{2+}$/H$^{+}$     &8.69($-$5) &1.13($-$4) &1.50($-$4)  &1.18($-$4) &2.40($-$4) &1.92($-$4)         \\
1550 &          C$^{3+}$/H$^{+}$     &6.55($-$5) &3.44($-$5) &1.67($-$5)  &    *      &1.33($-$4) &   *                \\
     &  \emph{icf}(C)                & 1.02      &1.28       &1.89        & 1.18      &1.06       &1.02              \\
     &           C/H                 &2.12($-$4) &2.37($-$4) &3.15($-$4)  &1.39($-$4) &4.35($-$4) &2.12($-$4)           \\
6548+6584 &     N$^{+}$/H$^{+}$      &1.66($-$7) &    *      &1.29($-$4)  &3.20($-$7) &1.27($-$5) &3.15($-$5)        \\
1750 &          N$^{2+}$/H$^{+}$     &1.29($-$5) &7.10($-$5) &9.95($-$5)  &2.20($-$5) &6.25($-$5) &1.93($-$4)         \\
57\,$\mu$m &    $^b$                 &  *        &      *    &8.68($-$5): &1.90($-$5):&2.92($-$5):&4.26($-$5):   \\
            &   $^c$                 &  *        &     *     &1.33($-$4): &4.13($-$5):&8.70($-$5):&2.76($-$4):    \\
1486 &          N$^{3+}$/H$^{+}$     &1.33($-$5) &2.71($-$5) &*           &    *      &6.77($-$6) &1.09($-$4)         \\
1240 &          N$^{4+}$/H$^{+}$     &  *        &1.22($-$5) &*           &    *      &2.35($-$5) &    *              \\
     &  \emph{icf}(N)                &1.09       & 1.31      &1.03        & 1.53      & 1.00      &  1.00                  \\
     &           N/H                 &2.87($-$5) &1.44($-$4) &2.36($-$4)  & 3.41($-$5)&1.05($-$4) &3.34($-$4)    \\ 
2470      &          O$^{+}$/H$^{+}$ &  *        &    *      &  *         &*          &  *        &4.24($-$5):   \\
3727      &          ...             &1.60($-$6) &    *      &3.03($-$4)  &2.54($-$6) &2.23($-$5) &4.19($-$5)           \\
7320+7330      &     ...             &1.02($-$6):&    *      &3.00($-$4): &4.12($-$6):&4.37($-$5):&4.24($-$5):       \\
1663      &          O$^{2+}$/H$^{+}$&5.72($-$5):&5.86($-$5):&2.35($-$4): &1.11($-$4):&2.34($-$4):&6.02($-$4):     \\
4931      &          ...             &1.04($-$4):&2.34($-$4):&2.71($-$4): &2.73($-$4):&2.57($-$4):&3.48($-$4):     \\
4959+5007 &          ...             &8.07($-$5) &9.71($-$5) &3.39($-$4)  &2.80($-$4) &2.90($-$4) &4.26($-$4)      \\
52+88\,$\mu$m &      $^b$            & *         & *         &2.94($-$4): &2.20($-$4):&1.64($-$4):&8.81($-$5):   \\
            &        $^c$            & *         &*          &3.48($-$4): &3.16($-$4):&3.91($-$4):&4.95($-$4):  \\
1401 &          O$^{3+}$/H$^{+}$     &2.86($-$4) &8.68($-$5) &   *        &    *      &3.51($-$4) &1.55($-$4)        \\
     &  \emph{icf}(O)                &1.24       & 1.34      &1.02        & 1.17      & 1.10      &  1.00           \\ 
     &           O/H                 &4.55($-$4) &2.46($-$4) &6.55($-$4)  & 3.31($-$4)&7.28($-$4) &6.23($-$4)             \\ 
3868+3967 &    Ne$^{2+}$/H$^{+}$     &1.59($-$5) &3.36($-$5) &1.58($-$4)  &6.61($-$5) &2.03($-$5) &1.26($-$4)    \\
15.5\,$\mu$m&   $^c$                 & *         &3.75($-$5):&2.41($-$5): &3.71($-$5):&5.75($-$5):&9.61($-$5):   \\
1601 &          Ne$^{3+}$/H$^{+}$    &5.35($-$5) &  *        &   *        &  *        &5.10($-$5) &7.35($-$5)          \\
2423 &          ...                  & *         &2.03($-$5) &   *        &  *        & *         &1.04($-$5):         \\
4725 &          ...                  &4.47($-$5) &   *       &   *        &9.18($-$6):&5.95($-$5) &   *               \\
1574 &          Ne$^{4+}$/H$^{+}$    &3.85($-$6) &9.38($-$5):&   *        &  *        & *         &1.60($-$3):     \\
3426 &          ...                  &   *       &5.59($-$5) &   *        &   *       &1.80($-$5) &  *          \\
     &  \emph{icf}(Ne)               & 1.00      & 1.00      &1.96        &1.18       &  1.00     &  1.00       \\
     &           Ne/H                &6.89($-$5) & 1.10($-$4)&3.10($-$4)  &7.80($-$5) &9.36($-$5) &2.00($-$4)    \\
4069+4076  &    S$^{+}$/H$^{+}$      &  *        &   *       &2.43($-$6): &3.37($-$9):&2.89($-$7):&1.30($-$6):   \\
6716+6730 &     ...                  &1.50($-$8) &   *       &3.21($-$6)  &1.45($-$8) &2.83($-$7) &1.24($-$6)     \\
6312 &          S$^{2+}$/H$^{+}$     &7.92($-$7) &   *       &6.80($-$6)  &6.62($-$7) & 1.93($-$6)&1.05($-$5)    \\
     &  \emph{icf}(S)                &4.57       &   *       &1.06        & 3.52      &  2.24     & 1.74           \\
     &           S/H                 &3.69($-$6) &   *       &1.06($-$5)  &2.38($-$6) &4.96($-$6) &2.04($-$5)            \\
5517+5537 &     Cl$^{2+}$/{\hp}      &1.76($-$8) &  *        &1.48($-$7)  & 2.44($-$8)&5.02($-$8) &1.31($-$7)     \\
     &  \emph{icf}(Cl)               &4.66       &  *        &1.56        &  3.59     & 2.57      & 1.95        \\
     &          Cl/H                 &8.20($-$8) &  *        &2.31($-$7)  &8.77($-$8) &1.29($-$7) &2.55($-$7)    \\
7135 &          Ar$^{2+}$/H$^{+}$    &3.39($-$7) &   *       &2.37($-$6)  &4.13($-$7) &8.87($-$7) &3.00($-$6)         \\
4711+4740 &     Ar$^{3+}$/H$^{+}$    &7.82($-$7) &5.12($-$7) &6.12($-$8)  &5.48($-$7) &5.88($-$7) &1.49($-$7)         \\
7005 &          Ar$^{4+}$/H$^{+}$    &2.13($-$7) &   *       &*           &7.16e$-$9  &1.15($-$7) &  *                \\
     &  \emph{icf}(Ar)               &1.00       &   *       &2.21        &  1.01     & 1.08      & 1.16           \\
     &           Ar/H                &1.33($-$6) &    *      &5.37($-$6)  &9.79($-$7) &1.72($-$6) &3.65($-$6)            \\
\noalign{\vskip3pt} \noalign{\hrule} \noalign{\vskip3pt}
\end{tabular}
\begin{description}
\item[$^a$] Values followed by `:' have not been used in order to derive total abundances.
\item[$^b$] Derived using the N$_e$ from the [O~{\sc iii}] 52~$\mu$m/88~$\mu$m line ratio.
\item[$^c$] Derived using the mean of the N$_e$'s obtained from the optical
[Cl~{\sc iii}] and [Ar~{\sc iv}] line ratios.
\end{description}
\end{minipage}
\end{table*}


\setcounter{table}{8}
\begin{table*}
\centering
\begin{minipage}{125mm}
\caption{Helium abundances derived from optical recombination lines and CEL
O, N, C, Ne, Ar, S and Cl abundances relative to hydrogen in Galactic PNe.$^a$}
\begin{tabular}{l@{\hspace{3mm}}l@{\hspace{3mm}}c@{\hspace{3mm}}c@{\hspace{3mm}}c@{\hspace{3mm}}c@{\hspace{3mm}}c@{\hspace{3mm}}c@{\hspace{3mm}}}
\noalign{\vskip3pt} \noalign{\hrule} \noalign{\vskip3pt}
$\lambda_0$\,({\AA})     &       &NGC\,5882  &NGC\,6302   &NGC\,6818  &IC\,4191     &IC\,4406   &My\,Cn\,18      \\
\noalign{\vskip3pt} \noalign{\hrule} \noalign{\vskip3pt}
4471 &  He$^{+}$/H$^{+}$         &0.1102     &0.0850      &0.0488     &0.0971       &0.1163     &   0.1057                  \\
5876 &  He$^{+}$/H$^{+}$         &0.1064     &0.0775      &0.0482     &0.1126       &0.1180     &   0.0984                  \\
6678 &  He$^{+}$/H$^{+}$         &0.1036     &0.0118      &0.0445     &0.1100       &0.1187     &  0.0862                   \\
Avg. &  He$^{+}$/H$^{+}$         &0.1066     &0.0658      &0.0476     &0.1090       &0.1178     &0.0974               \\
4686 &  He$^{2+}$/H$^{+}$        &0.0022     &0.0696      &0.0510     &0.0110       &0.0067     &0.0004               \\
     &  He/H                     &0.109      &0.135       &0.099      &0.120        &0.124      &0.098            \\
2326 &          C$^{+}$/H$^{+}$  &  *        &  *         &2.63($-$5) &  *          &1.34($-$4) & *                    \\
1908 &          C$^{2+}$/H$^{+}$ &1.29($-$4) &3.08($-$5)  &2.04($-$4) & *           &2.11($-$4) & *                        \\
1550 &          C$^{3+}$/H$^{+}$ &   *       &1.43($-$5)  &2.21($-$5) & *           &1.81($-$5) & *                        \\
     &  \emph{icf}(C)            &  1.17     & 1.72       &1.03       & *           &1.00       &   *                  \\
     &           C/H             &1.51($-$4) &7.74($-$5)  & 2.60($-$4)& *           &3.63($-$4) &    *                        \\
6548+6584 &     N$^{+}$/H$^{+}$  &3.06($-$6) &4.48($-$5)  &8.29($-$6) &9.64($-$6)   &5.90($-$5) &8.11($-$5)           \\
1750 &          N$^{2+}$/H$^{+}$ &1.10($-$4) &1.03($-$4)  &3.44($-$5) &*            &1.10($-$4) & *                     \\
57\,$\mu$m &    $^b$             &5.08($-$5):&5.20($-$5): & *         &2.45($-$5)   &9.05($-$5):& *         \\
           &    $^c$             &1.03($-$4):&3.13($-$4): &*          &1.34($-$4):  &2.63($-$4):&*     \\
1486 &          N$^{3+}$/H$^{+}$ &    *      &7.80($-$5)  &6.06($-$6) &   *         &  *        & *                       \\
1240 &          N$^{4+}$/H$^{+}$ &    *      &9.45($-$5)  &6.24($-$6) & *           &  *        & *                           \\
     &  \emph{icf}(N)            &   1.35    & 1.00       &   1.00    & 1.14        & 1.25      &2.71                     \\
     &           N/H             &1.52($-$4) &3.34($-$4)  &5.50($-$5) &3.89($-$5)   & 2.11($-$4)& 2.20($-$4)             \\
2470 &          O$^{+}$/H$^{+}$  &           &1.67($-$5): &2.37($-$5):&  *          &2.03($-$4):&  *                        \\
3727 &          ...              &1.31($-$5) &8.42($-$6)  &2.28($-$5) &2.26($-$5)   &1.79($-$4) &2.08($-$4)          \\
7320+7730 &     ...              &2.17($-$5):&2.03($-$5): &2.37($-$5):&2.81($-$5):  &1.70($-$4):&4.78($-$4):         \\
1663 &          O$^{2+}$/H$^{+}$ &4.35($-$4):&9.35($-$5): &1.51($-$4):& *           &3.86($-$4):&  *                   \\
4931 &          ...              &4.38($-$4):&2.66($-$4): &1.68($-$4):&4.05($-$4):  &3.67($-$4):&3.76($-$4):             \\
4959+5007 &     ...              &4.72($-$4) &9.18($-$5)  & 2.52($-$4)&5.46($-$4)   &3.80($-$4) &3.54($-$4)         \\
52+88\,$\mu$m &  $^b$            &2.70($-$4):&4.20($-$5): & *         &1.90($-$4):  &3.90($-$4):&*     \\
      &          $^c$            &4.70($-$4):&9.90($-$5): &*          &8.36($-$4):  &7.78($-$4):&*  \\
1401 &          O$^{3+}$/H$^{+}$ &   *       &8.22($-$5)  &1.84($-$4) &*            &  *        & *                   \\
     &  \emph{icf}(O)            & 1.01      &  1.37      &1.11       & 1.07        & 1.04      & 1.00             \\
     &           O/H             & 4.90($-$4)&2.50($-$4)  & 5.08($-$4)&6.08($-$4)   &5.81($-$4) & 5.64($-$4)             \\
3868+3967 &     Ne$^{2+}$/H$^{+}$&1.31($-$4) &2.65($-$5)  &5.57($-$5) &1.47($-$4)   &1.39($-$4) &7.74($-$5)                   \\
15.5\,$\mu$m&    $^c$            &1.47($-$4):&1.81($-$5): &5.18($-$5):& *           & *         &*       \\
1601 &          Ne$^{3+}$/H$^{+}$&   *       &3.53($-$5)  &5.00($-$5) &*            &  *        &   *                      \\
2423 &          ...              &   *       &7.59($-$5): &1.45($-$5):& *           &1.01($-$5):& *                   \\
4724 &          ...              &   *       &2.37($-$5)  &5.55($-$5) &7.21($-$5):  &  *        & *                    \\
1574 &          Ne$^{4+}$/H$^{+}$&   *       &  *         &2.05($-$4):&*            &  *        & *                       \\
3426 &          ...              &   *       &2.00($-$5)  &9.57($-$6) &  *          &           &  *\\
     &  \emph{icf}(Ne)           &  1.04     &1.00        & 1.00      & 1.11        & 1.53      & 1.59                    \\
     &           Ne/H            &1.36($-$4) &7.60($-$5)  & 1.18($-$4)& 1.64($-$4)  &2.13($-$4) & 1.23($-$4)               \\
4069+4076 &     S$^{+}$/H$^{+}$  &8.91($-$8):&8.20($-$7): &1.17($-$7):&3.45($-$7):  &4.20($-$7):&1.43($-$6):                       \\
6716+6730 &     ...              &1.44($-$7) &1.04($-$6)  &2.73($-$7) &5.59($-$7)   &4.77($-$7) &8.64($-$7)                        \\
6312 &          S$^{2+}$/H$^{+}$ &3.50($-$6) &1.57($-$6)  & 1.54($-$6)&5.49($-$6)   &1.38($-$6) &1.35($-$5)                  \\
     &  \emph{icf}(S)            &2.30       & 2.14       & 2.01      & 2.10        &1.14       &  1.20                     \\
     &           S/H             &8.38($-$6) &5.59($-$6)  & 3.64($-$6)&1.27($-$5)   &2.12($-$6) & 4.43                   \\
5517+5537 &     Cl$^{2+}$/{\hp}  &7.59($-$8) &2.76($-$8)  &5.24($-$8) &1.02($-$7)   &9.37($-$8) &2.09($-$7)        \\
     & \emph{icf}(Cl)            & 2.40      &3.56        &2.37       &2.31         &1.54       &1.17           \\
     &          Cl/H             &1.82($-$7) &9.83($-$8)  &1.24($-$7) &2.36($-$7)   &1.44($-$7) &2.45($-$7)        \\
7135 &          Ar$^{2+}$/H$^{+}$&1.37($-$6) &6.51($-$7)  &8.41($-$8) & 1.66($-$6)  &1.64($-$6) &1.88($-$6)                      \\
4711+4740 &     Ar$^{3+}$/H$^{+}$&6.31($-$7) &7.75($-$7)  &6.08($-$7) & 5.99($-$7)  &1.80($-$7) & *                         \\
7005 &          Ar$^{4+}$/H$^{+}$&   *       &4.67($-$7)  &9.57($-$8) &7.37($-$8)   & *         & *                        \\
     &  \emph{icf}(Ar)           &  1.02     &1.16        &1.11       & 1.33        &1.39       & *                       \\
     &           Ar/H            &2.04($-$6) &2.19($-$6)  &8.73($-$7) & 3.10($-$6)  &2.53($-$6) &*                         \\
\noalign{\vskip3pt} \noalign{\hrule} \noalign{\vskip3pt}
\end{tabular}
\begin{description}
\item[$^a$] Values followed by `:' have not been used in order to derive total
abundances;
\item[$^b$] Derived using the {\eld} from the [O~{\sc iii}] 52~$\mu$m/88~$\mu$m line
ratio;
\item[$^c$] Derived using the mean of the {\eld}'s obtained from the optical
[Cl~{\sc iii}] and [Ar~{\sc iv}] line ratios.
\end{description}
\end{minipage}
\end{table*}


\setcounter{table}{9}
\begin{table}
\centering
\begin{minipage}{85mm}
\caption{Ionic and total elemental abundances relative to hydrogen for helium
derived from ORLs and for heavy elements derived from CELs in Magellanic Cloud
PNe.$^a$} \footnotesize
\begin{tabular}{l@{\hspace{3mm}}l@{\hspace{3mm}}c@{\hspace{3mm}}c@{\hspace{3mm}}c@{\hspace{3mm}}}
\noalign{\vskip3pt} \noalign{\hrule} \noalign{\vskip3pt}
$\lambda_0$\,({\AA})   &               &SMC\,N87   &LMC\,N66   &LMC\,N141 \\
 \noalign{\vskip3pt} \noalign{\hrule}  \noalign{\vskip3pt}

4471 &  He$^{+}$/H$^{+}$               &  0.0965  &  0.0572  & 0.1058   \\
5876 &  He$^{+}$/H$^{+}$               &    *     &  0.0375  &    *     \\
6678 &  He$^{+}$/H$^{+}$                & 0.0981   & 0.0435   &0.1119       \\
Avg. &  He$^{+}$/H$^{+}$                &0.0972    &0.0440    &0.1085       \\
4686 &  He$^{2+}$/H$^{+}$               &0.0000    &0.0607    &0.0003       \\
     &  He/H                            &0.097     &0.105     &0.109        \\
2326 &          C$^{+}$/H$^{+}$         &3.59($-$5)   &*      &3.02($-$5)          \\
1908 &          C$^{2+}$/H$^{+}$        &3.22($-$4)   &5.92($-$6)&1.49($-$4)      \\
1550 &          C$^{3+}$/H$^{+}$        &1.92($-$5)   &7.77($-$6)&2.04($-$5)      \\
     &  \emph{icf}(C)                    &1.00      &2.40  & 1.00            \\
     &           C/H                    &3.77($-$4)   &3.29($-$5)  & 2.00($-$4)      \\
6548+6584 &     N$^{+}$/H$^{+}$         &6.10($-$7)   &7.44($-$6)   &2.03($-$6)      \\
1750 &          N$^{2+}$/H$^{+}$        &*         &1.84($-$5)   &*            \\
1486 &          N$^{3+}$/H$^{+}$        & *        &4.46($-$5)   &*            \\
1240 &          N$^{4+}$/H$^{+}$         &*        &2.70($-$5)   &2.81($-$5):      \\
     &  \emph{icf}(N)                    & 58.7     &1.00    & 44.0        \\
     &           N/H                    &3.58($-$5)   &9.74($-$5)&8.93($-$5)         \\
3727 &          O$^{+}$/H$^{+}$         &1.84($-$6)   &5.95($-$6)   &4.48($-$6)      \\
7320+7330 &     ...                     &1.38($-$5):  &2.15($-$5):  &3.25($-$5):     \\
1663 &          O$^{2+}$/H$^{+}$        &*         &4.58($-$5):  &*            \\
4931 &          ...                     &7.94($-$5):  &  *      &1.04($-$4):         \\
4959 &          ...                     &1.06($-$4)   &7.81($-$5)   &1.93($-$4)      \\
1401 &          O$^{3+}$/H$^{+}$        &*         & *    &*                \\
     &  \emph{icf}(O)                   & 1.00     & 3.77 & 1.00            \\
     &           O/H                    &1.08($-$4)   &3.17($-$4)&1.97($-$4)         \\
3868+3967 &     Ne$^{2+}$/H$^{+}$       &1.04($-$5)   &1.74($-$5)   &2.35($-$5)      \\
1601 &          Ne$^{3+}$/H$^{+}$       &*         &*       &*                \\
2423 &          ...                     & *        &1.39($-$5)  &*              \\
4724 &          ...                     & *        &1.90($-$5)  &*            \\
1574 &          Ne$^{4+}$/H$^{+}$       & *        &*        &*            \\
3426 &          ...                     & *        &8.24($-$6)  & *           \\
     &  \emph{icf}(Ne)                   & 1.02     &1.00   &1.02            \\
     &           Ne/H                   & 1.06($-$5)  &4.21($-$5)   &2.40($-$5)           \\
4069 &          S$^{+}$/H$^{+}$         &4.10($-$8):  & *        &1.21($-$7):             \\
6716+6730 &     ...                     &2.85($-$8)   &2.43($-$7)   &1.48($-$7)          \\
6312 &          S$^{2+}$/H$^{+}$        &*         &7.69($-$7)   &1.50($-$6)                \\
     &  \emph{icf}(S)                    &*         &2.11         &2.47                 \\
     &           S/H                    &*         &4.30($-$6)       &4.06($-$6)           \\
7135 &          Ar$^{2+}$/H$^{+}$       &1.76($-$7)   &4.28($-$7)   &6.39($-$7)          \\
4711+4740 &          Ar$^{3+}$/H$^{+}$       &3.13($-$8)   &5.43($-$7)   &1.03($-$7)          \\
7005 &          Ar$^{4+}$/H$^{+}$       &*         &1.57($-$7)   &*                \\
     &  \emph{icf}(Ar)                   &1.02      &1.13      &1.02           \\
     &           Ar/H                   &2.11($-$7)   &1.27($-$6)   &7.59($-$7)         \\
\noalign{\vskip3pt} \noalign{\hrule} \noalign{\vskip3pt}
\end{tabular}
\begin{description}
\item[$^a$] Values followed by `:' have not been used in order to derive total abundances.
\end{description}
\end{minipage}
\end{table}

Abundances of neutral species were not derived; it is assumed that heavy
element neutral fractions are the same as hydrogen neutral fractions and
therefore that total elemental abundances relative to hydrogen can be obtained
using ionic fractions only. In Tables~9 and 10 the abundances for all observed
ionic species of C, N, O, Ne, S, Cl and Ar are presented, together with the
adopted ionization correction factors (ICFs) and total elemental abundances, as
derived using CELs only. The adopted ICF scheme is that of KB94; details on the
derivation of the total abundances for C, N and O, are discussed in the
Appendix.

The adopted O$^+$/H$^+$ fractions used to calculate the total oxygen abundances
were obtained from the $\lambda$3727 doublet only. The abundances derived from
the {\foii} $\lambda\lambda$7320, 7330 lines are generally higher, as is
evident from Tables~9 and 10; the reason could be that a larger fraction of
these line fluxes with respect to that of $\lambda$3727 can be due to
recombination excitation and/or the fact that the emission of
$\lambda\lambda$7320, 7330 is biased towards higher density regions (cf.
Section~4.2.1). However, the ionic fractions deduced from $\lambda$3727 may
represent upper limits only due to contributions from recombination.

In the case of {\opp} we adopted ionic fractions obtained from the nebular
{\foiii} $\lambda\lambda$4959, 5007 lines only, in order to derive total oxygen
abundances. For comparison, in Tables~9 and 10, we also present {\opp}/{\hp}
abundances derived from the O~{\sc iii}] $\lambda$1663 and {\foiii}
$\lambda$4931 lines. We further present {\opp}/{\hp} abundances derived from
the far-IR {\foiii} 52- and 88-$\mu$m fine-structure lines for eight nebulae,
using the \emph{ISO} LWS line fluxes published in Liu et al. (2001a); two
values per nebula are listed---for the first we used the (lower) electron
densities obtained from the 88\,$\mu$m/52\,$\mu$m line ratio and for the second
we used the (higher) {\eld}'s obtained from the  optical {\fariv} and {\fcliii}
density diagnostics (Table~6).

Similarly, we present {\npp}/{\hp} ionic fractions from both the N~{\sc iii}]
$\lambda1750$ line and from the far-IR {\fniii} 57-$\mu$m line (using the
\emph{ISO} LWS 57-$\mu$m line fluxes of Liu et al. 2001a); abundance
ratios from the latter line are tabulated for both the low and high
nebular electron density cases, just as for the {\foiii} far-IR lines.

Both the {\foiii} and {\fniii} far-IR lines originate from atomic levels that
have quite low critical densities,
{\crd}\,$\sim$\,(2--4)\,$\times$\,10$^{3}$\,{\cmt}, lower than the average
electron density of most PNe in our sample. The ionic abundances derived from
these lines are therefore acutely sensitive to our assumption of the actual
density of the emitting medium. For example, adopting a density significantly
higher than the {\crd} of the collisional line will result in a proportional
increase of the deduced abundance, since for the same flux to be emitted the
presence of more ions is required, otherwise the emission line would be
quenched. On the other hand, the nebular {\foiii} $\lambda\lambda$4959, 5007
lines have much higher critical densities,
{\crd}\,$\sim$\,6.9\,$\times$\,10$^{5}$\,{\cmt}, so that the abundances deduced
from them are much less sensitive to the adopted nebular density; the same is
true for the N~{\sc iii}] $\lambda$1750 line which has
{\crd}\,$\sim$\,11\,$\times$\,10$^{9}$\,{\cmt}.

\setcounter{figure}{2}
\begin{figure*}
\begin{center} \epsfig{file=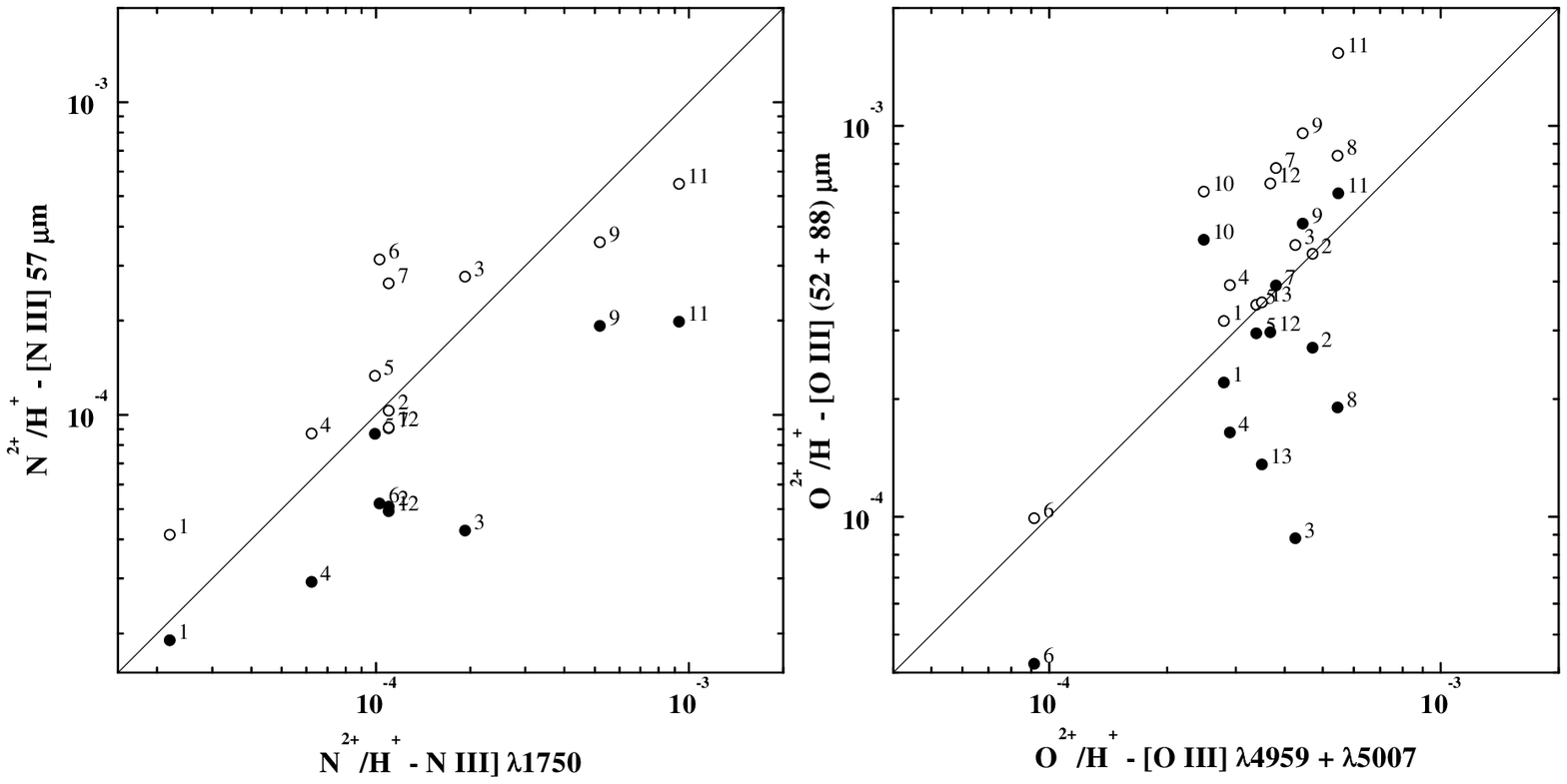, width=18. cm, clip=}
\caption{Comparison of {\npp}/{\hp} abundance ratios derived from the N~{\sc
iii}] $\lambda$1750 and {\fniii} 57\,$\mu$m lines, and {\opp}/{\hp} ratios
derived from the {\foiii} $\lambda\lambda$4959, 5007 and 52, 88\,$\mu$m lines;
solid and open circles denote abundances derived from adopting `low' far-IR
{\foiii} 88\,$\mu$m/52\,$\mu$m electron densities and `high' {\fcliii},
{\fariv} electron densities, respectively; the data labelling refers to the
designated PN numbers of Table~11, Col. 2.}
\end{center}
\end{figure*}

Thus {\opp}/{\hp} and {\npp}/{\hp} ionic abundances derived from far-IR lines
are expected to be underestimated in the presence of density variations in
nebulae (Rubin 1989; Liu et al. 2001a). The likelihood of systematic bias in
abundance determinations from CELs becomes less when lines with fairly high
critical densities are used, as long as {\eld} values in the nebula are well
below those values. These expectations are confirmed by our observations.
Table~6 compares the electron densities derived from the 88\,$\mu$m/52\,$\mu$m
line ratio with those derived from the higher critical density {\fcliii}, and
{\fariv} ratios, for the eight PNe in common between our sample and Liu et
al.'s (2001a) \emph{ISO} LWS sample, and shows that on average the latter
diagnostics yield {\eld}'s about a factor of 6 higher than the far-IR ratio,
confirming the existence of density inhomogeneities in these nebulae, in
agreement with Liu et al.'s conclusions. In accord with the predictions of
Rubin, the bias in the ionic {\npp} and {\opp} abundances derived from the
far-IR lines becomes \emph{less} when the higher electron densities from the
{\fcliii} and {\fariv} are adopted -- the values returned are then in good
agreement, in most cases, with those derived from the {\fniii} $\lambda$1750
and {\foiii} $\lambda\lambda$4959, 5007 lines (cf. Table~9). For instance, in
the case of NGC\,5882 the `high' density IR {\opp} abundance is within 1\,per
cent of the optical CEL value, while the `high' density IR {\npp} abundance is
within 7\,per cent of the UV CEL value; in the case of NGC\,3918, the UV {\npp}
and optical {\opp} abundances agree to within 8 and 5\,per cent with the
corresponding values that are \emph{half-way} between the listed IR abundances.
A notable exception is IC\,4406, for which the {\opp}/{\hp} abundance derived
from the far-IR lines in the low density case agrees with the corresponding
value derived from the optical lines. In Fig.\,~3 we plot abundance ratios of
{\npp}/{\hp} and {\opp}/{\hp} derived from UV and optical lines respectively
and compare them with those derived from far-IR transitions. Regarding the
abundances from the IR lines, we plot using different symbols both those
resulting from adopting the low 88\,$\mu$m/52\,$\mu$m ratio {\eld}'s as well as
those derived from adopting the higher {\fcliii}, {\fariv} {\eld}'s. We see
that for the majority of nebulae very satisfactory agreement is found amongst
the UV, and optical abundances on the one hand and the far-IR abundances on the
other, after accounting for the modest density variations within the nebular
volumes, exposed from the disparity between the IR and optical density
diagnostic line ratios. In general, the adoption of electron densities just a
little below those diagnosed by the {\fcliii} and {\fariv} ratios would lead to
agreement between the abundances deduced from the infrared FS lines of {\foiii}
and {\fniii} and those derived from the optical or UV lines of the same ions.
These results are of importance in the context of the discussion about the
existence of temperature fluctuations in nebulae and their potential impact on
abundances derived from CELs (cf. Section~4.2.3).

We derived {\nepp}/{\hp} abundances from the optical {\fneiii}
$\lambda\lambda$3868, 3967 lines for all 15 PNe. For 9 of them we also
determined this abundance ratio using the \emph{IRAS} {\fneiii} 15.5-$\mu$m
line fluxes of Pottasch et al. (1984). For the 15.5-$\mu$m line, {\crd} =
2\,$\times$\,10$^{5}$\,{\cmt}, and the resulting ionic abundances do not depend
on the adopted {\eld}, for the range of nebular densities presented in Table~6.
Inspection of Table~9 shows quite good agreement between the optical and
infrared {\nepp}/{\hp} abundances. As with other ions, the abundances from the
optical CELs are those adopted.

In all these cases where IR collisionally excited lines were analyzed, we
adopted the {\elt}(\oiii)'s for abundance determinations. IR CELs have small
excitation energies, {\exe}\,$<$\,1000\,K, much smaller than those of UV and
optical forbidden lines. Thus their emissivities have only a weak dependence on
the adopted nebular electron temperature, very similar to those of {\hi} Balmer
lines. Their intensities relative to {\Hb} are therefore virtually insensitive
to the assumed temperature, unless the emitting medium has
{\elt}\,$\ll$\,1000\,K; the electron temperatures of all nebulae in our sample,
derived from CEL intensity ratios, are consistently much higher than that. On
the other hand, at the temperatures implied by the {\oii} ORLs (cf. Table~7 and
Paper~II) the far-IR CELs would be expected to be suppressed to various degrees
depending on the exact temperature and density of the emitting material.

{\neppp}/{\hp} ionic fractions were derived from the {\fneiv}
$\lambda\lambda$4724, 4726 and $\lambda$1601 lines, which usually showed good
agreement with each other. For the {\nefp}/{\hp} ratio, whenever available the
values obtained from [Ne~{\sc v}] $\lambda$3426 were preferred over those from
the Ne~{\sc v} $\lambda$1574 line.

We adopted the S$^+$/H$^+$ abundance derived from the {\fsii}
$\lambda\lambda$6716, 6731 doublet, rather than from the transauroral
$\lambda\lambda$4068, 4076 lines, since the latter are potentially affected by
recombination processes and density effects (the $^{\rm 2}D$ and $^{\rm 2}P$
atomic levels of S$^+$, from which the $\lambda\lambda$6716, 6731 and
$\lambda\lambda$4068, 4076 lines respectively arise, are directly analogous to
the levels of O$^+$ from which the {\foii} nebular and auroral lines originate;
especially in terms of critical densities).

\subsection{Abundances -- Summary}

Table~11 presents total elemental abundances by number, expressed in units of
log\,(X/H) + 12.0, for X = He, C, N, O, Ne, S, Cl and Ar, for the 15 PNe
analyzed in this work, plus the Galactic nebulae NGC\,7009 (LSBC), NGC\,6153
(Liu et al. 2000), and M\,1-42 and M\,2-36 (Liu et al. 2001b) which were
previously observed in the context of our ongoing programme. Peimbert \&
Torres-Peimbert (1983) defined Type~I PNe, believed to represent the high mass
end of the PN distribution, as those having He/H $\ge$ 0.125 and N/O $\ge$ 0.5
by number. KB94 refined the Type~I classification to refer to those PNe which
have N/O ratios larger than the original (C + N)/O ratios of the parent galaxy
ISMs out of which they formed, implying that any nitrogen in excess of this
value must be primary in origin, e.g. via hot bottom burning of carbon brought
up by the third dredge-up. For the Milky Way, the KB94 Type~I criterion
translates into N/O $\ge$ 0.74.  Using this, we classify NGC\,2440, NGC\,6302,
NGC\,7009 and M\,1-42 as Type~I. Table~11 also presents mean elemental
abundances for the twelve non-Type~I and four Type~I Galactic PNe listed there
and compares them with the mean non-Type~I and Type~I PN abundances of KB94, as
well as with solar and M\,42 (Orion Nebula) abundances. The mean N, O, Ne and S
abundances for the twelve non-Type~I Galactic PNe are slightly larger than the
means found by KB94 for their larger sample of Galactic PNe. However, this may
be because M\,2-36, which has unusually large abundances for all elements,
skews the mean values for our sample.

In Paper~II we will use the much weaker heavy element optical recombination
lines (ORLs) measured in the spectra reported here to derive ionic abundances
and electron temperatures that can be compared with those derived in this paper
from the CELs, in order to investigate whether our data replicate, and can
throw light on, the significant discrepancies that have previously been found
between CEL and ORL ionic abundances measured for a number of other planetary
nebulae.

\begin{table}
\begin{center}
\caption{Elemental abundances relative to H derived from
CELs, except for the helium abundances which are from ORLs, in units where log\,$N$(H) = 12.0.$^a$}
\begin{tabular}{l@{\hspace{0.4mm}}r@{\hspace{1mm}}c@{\hspace{2.mm}}c@{\hspace{2.mm}}c@{\hspace{2.mm}}c@{\hspace{2.mm}}c@{\hspace{2.mm}}c@{\hspace{2.mm}}c@{\hspace{2.mm}}c@{\hspace{2.mm}}}
\noalign{\vskip3pt}\noalign{\hrule}\noalign{\vskip3pt}
 PN          &No.  &He      &C        &N       &O     &Ne     &S        &Cl       &Ar \\
\noalign{\vskip3pt}\noalign{\hrule}\noalign{\vskip3pt}
NGC\,2022    &     &11.04   &8.33     &7.46    &8.66  &7.84   &6.57     &4.91     &6.12     \\
NGC\,2440 (I)&     &11.08   &8.37     &8.26    &8.39  &8.04   & *       & *       & *       \\
NGC\,3132    &(5)  &11.08   &8.50     &8.37    &8.82  &8.49   &7.03     &5.36     &6.73      \\
NGC\,3242    &(1)  &11.00   &8.14     &7.53    &8.52  &7.89   &6.38     &4.94     &5.99    \\
NGC\,3918    &(4)  &11.00   &8.64     &8.02    &8.86  &7.97   &6.70     &5.11     &6.24     \\
NGC\,5315    &(3)  &11.08   &8.33     &8.52    &8.79  &8.30   &7.31     &5.41     &6.56     \\
NGC\,5882    &(2)  &11.04   &8.18     &8.18    &8.67  &8.13   &6.92     &5.26     &6.31     \\
NGC\,6153$^b$ &(9) &11.14   &8.44     &8.36    &8.70  &8.23   &7.21     &5.62     &6.43    \\
NGC\,6302 (I)&(6)  &11.13   &7.89     &8.52    &8.40  &7.88   &6.75     &4.99     &6.34    \\
NGC\,6818    &     &11.00   &8.41     &7.74    &8.71  &8.07   &6.56     &5.09     &5.94    \\
NGC\,7009$^c$ (I)&(12)&11.04   &8.66     &8.50    &8.61  &8.24   &6.98     & *       &6.27    \\
IC\,4191     &(8)  &11.08   &*        &7.59    &8.78  &8.21   &7.10     &5.37     &6.49   \\
IC\,4406     &(7)  &11.09   &8.56     &8.32    &8.76  &8.33   &6.33     &5.16     &6.40    \\
My\,Cn\,18   &     &10.99   &*        &8.34    &8.75  &8.09   &7.24     &5.39     & *         \\
\noalign{\vskip2pt}
M\,1-42 $^d$ (I)&(10)&11.17  &7.80     &8.68    &8.63  &8.12   &7.08     &5.26     &6.56    \\
M\,2-36 $^d$   &(11)&11.13  &8.73     &8.42    &8.85  &8.57   &7.47     &5.42     &6.66    \\
\noalign{\vskip2pt}
Mean non-Type~I&   &11.06   &8.46     &8.30    &8.75  &8.23   &7.04     &5.30     &6.37     \\
Mean Type~I    &   &11.11   &8.32     &8.51    &8.52  &8.09   &6.96     &5.15     &6.41     \\
\noalign{\vskip2pt}
KB94 non Type-I$^e$  & &11.05        &8.81     &8.14    &8.69  &8.10    &6.91    &*        &6.38  \\
KB94 Type-I$^e$      & &11.11         &8.48     &8.72    &8.65  &8.09    &6.91    &*        &6.42\\
Solar$^f$       &  &10.99    &8.39     &7.97    &8.69  &8.09   &7.21    &5.50     &6.56  \\
M\,42$^g$       &  &10.99    &8.53     &7.78    &8.52  &7.89   &7.17    &5.33     &6.80  \\
\noalign{\vskip2pt}
SMC N87       &   &10.99    &8.58     &7.55    &8.03  &7.03   & *       & *       &5.32  \\
LMC N66       &   &11.02    &7.52     &7.99    &8.50  &7.62   &6.63     & *       &6.10      \\
LMC N141      &   &11.04    &8.30     &7.95    &8.29  &7.38   &6.61     & *       &5.88       \\
\noalign{\vskip3pt} \noalign{\hrule} \noalign{\vskip3pt}
\end{tabular}
\begin{description}
\item[$^a$] Values are for the whole nebulae, except NGC\,6302, 6644, 6572, 6818, 7009, My\,Cn\,18, IC\,4406, LMC~N66
and M\,42, where values are from fixed-slit spectra. A ``(I)'' after a PN's name indicates that it is a Type~I
nebula (see text).
\item[$^b$] From Liu et al. (2000).
\item[$^c$] From LSBC for He \& O; Henry, Kwitter \& Bates (2000) for C \& N; Luo, Liu \& Barlow (2001) for Ne; KB94 for S \& Ar.
\item[$^d$] From Liu et al. (2001b).
\item[$^e$] From KB94 mean abundances.
\item[$^f$] Solar photospheric abundances from Grevesse et al. (1996),
except C and O which are from Allende Prieto et al. (2001, 2002).
\item[$^g$] CEL abundances for M~42 from Esteban et al. (1998), except O, which is from
our unpublished ESO 1.52-m and AAT 3.9-m data, and C, which is from Rubin et al. (1991).
\end{description}
\end{center}
\end{table}

\vspace{7mm} \noindent {\bf Acknowledgments}

\noindent{YGT acknowledges the award of a Perren studentship.}

\begin{appendix}

\section{ICF method for the derivation of C, N, and O}

Generally, in order to calculate total elemental abundances we made use of the
\emph{icf} scheme of KB94, with some modifications in those cases \emph{only}
where our observations of heavy-element optical recombination lines (and the
resultant ORL abundances; Paper II) enabled us to account in a satisfactory
manner for missing ionic stages. Below we describe how we derived the
\emph{icf}s for C, N, and O from this method.

{\bf NGC\,2022}: In the case of carbon we use the standard KB94 correction
factor 1 + {\op}/{\opp} to account for missing {\cp}, while from the ratio
{\cfp}/{\cppp} as derived from ORLs and the CEL {\cppp} abundance we estimate
{\cfp}/{\hp} = 5.53\,$\times\,10^{-5}$ [thus icf(C) = 1.02]. For N, we use the
ratio {\nfp}/{\nppp} as derived from ORLs and the CEL {\nppp} abundance to
estimate {\nfp}/{\hp} = 2.35\,$\times\,10^{-6}$ [thus icf(N) = 1.09]. Similarly
for O, we estimate {\ofp}/{\hp} = 8.69\,$\times\,10^{-5}$ from the
{\ofp}/{\oppp} ratio derived from ORLs and the CEL {\oppp} abundance [thus
icf(O) = 1.24].

{\bf NGC\,2440}: We treat C exactly as in the case of NGC\,2022, using the
standard correction for {\cp} and estimating {\cfp}/{\hp} =
3.77\,$\times\,10^{-5}$ with the help of the ORL {\cfp}/{\cppp} ratio and the
CEL {\cppp} abundance [thus icf(C) = 1.89]. Our C abundance is only 2\,per cent
higher than the one found if we follow the standard \emph{icf} scheme based on
nitrogen. For nitrogen, we make use of the CEL {\np}/{\npp} ratio given by KB94
and our own {\npp} abundance to estimate {\np}/{\hp} = 3.37\,$\times\,10^{-5}$
[thus icf(N) = 1.31]. For oxygen, since our spectral coverage of the nebula did
not allow the derivation of an {\op} abundance, we made use of the CEL ratio
{\op}/{\opp} from Liu \& Danziger (1993; from their 270 deg. slit), along
with our {\opp}
abundance to estimate {\op}/{\hp} = 2.76\,$\times\,10^{-5}$. We also estimate
{\ofp}/{\hp} = 3.43\,$\times\,10^{-5}$ using our ORL ratio {\ofp}/{\oppp} and
the CEL {\oppp}/H$^+$ abundance ratio [thus icf(O) = 1.34].

{\bf NGC\,3132}: The standard schemes were used in the case of C and O, while
for nitrogen we use our ORL {\nppp}/{\npp} ratio and the CEL {\npp} abundance
to estimate {\nppp}/{\hp} = 7.22\,$\times\,10^{-6}$ [thus icf(N) = 1.03].

{\bf NGC\,3242}: For C and O we used the standard scheme, while for nitrogen
using our ORL {\nppp}/{\npp} ratio and the CEL {\npp} abundance we find
{\nppp}/{\hp} = 1.18\,$\times\,10^{-5}$ [thus icf(N) = 1.53].

{\bf NGC\,3918}: For N and O we follow the standard scheme, while for carbon we
use the ratio {\cfp}/{\cppp} as derived from ORLs and the CEL {\cppp} abundance
to estimate {\cfp}/{\hp} = 2.62\,$\times\,10^{-5}$ [thus icf(C) = 1.06].

{\bf NGC\,5315}: For N and O we follow the standard scheme, while for carbon we
use our ORL {\cppp}/{\cpp} ratio and the CEL {\cpp} abundance to estimate
{\cppp}/{\hp} = 4.61\,$\times\,10^{-6}$ [thus icf(C) = 1.02].

{\bf NGC\,5882}: For O we used the standard scheme; for carbon we estimate
{\cppp}/{\hp} = 1.81\,$\times\,10^{-5}$ from the ORL {\cppp}/{\cpp} ratio and
the CEL {\cpp} abundance [thus icf(C) = 1.17]. For nitrogen we use our ORL
{\nppp}/{\npp} ratio and the CEL {\npp} abundance to estimate {\nppp}/{\hp} =
3.88\,$\times\,10^{-6}$ [thus icf(N) = 1.35].

{\bf NGC\,6818}: For carbon we used the ratio {\cfp}/{\cppp} as derived from
ORLs and the CEL {\cppp} abundance to estimate {\cfp}/{\hp} =
7.20\,$\times\,10^{-6}$ [thus icf(C) = 1.03]. For N we used the standard
scheme, while for oxygen we use our ORL ratio {\ofp}/{\oppp} and the CEL
{\oppp} abundance to estimate {\ofp}/{\hp} = 4.97\,$\times\,10^{-5}$ [thus
icf(O) = 1.11].

{\bf IC\,4191}: We used the standard scheme for C and O, whereas for nitrogen
we used our ORL {\nppp}/{\npp} ratio and the CEL {\npp} abundance to estimate
{\nppp}/{\hp} = 4.80\,$\times\,10^{-6}$ [thus icf(N) = 1.14].

{\bf IC\,4406}: We used the standard scheme for C and O, whereas for nitrogen
we used our ORL {\nppp}/{\npp} ratio and the CEL {\npp} abundance to estimate
{\nppp}/{\hp} = 4.23\,$\times\,10^{-5}$ [thus icf(N) = 1.25]. \\

Finally, for NGC\,6302, My\,Cn\,18, SMC N87, LMC N66 and LMC N141 we used the
standard scheme throughout.

\end{appendix}

\end{document}